\documentclass[]{tCPH2e}

\def\apj{ApJ\,}
\def\apjl{ApJL\,}
\def\apjs{ApJS\,}
\def\aj{AJ\,}
\def\mnras{MNRAS\,}
\def\araa{ARA\&A\,}
\def\aapr{A\&AR\,}
\def\prd{PRD\,}
\def\nat{Nature\,}
\def\aap{A\&A\,}
\def\ssr{Space Science Reviews\,}
\def\physrep{Phys. Rep.\,}
\usepackage{sidecap}

\begin{document}
\doi{10.1080/0010751YYxxxxxxxx}
 \issn{1366-5812}
\issnp{0010-7514}

\jvol{00} \jnum{00} \jyear{2010} 

\markboth{Treu \& Ellis}{Contemporary Physics}


\title{{\itshape Gravitational Lensing - Einstein's Unfinished Symphony
 }}

\author{Tommaso Treu$^{a}$$^{\ast}$\thanks{$^\ast$Corresponding author. Email: tt@astro.ucla.edu
\vspace{6pt}} and Richard S. Ellis$^{b}$\\\vspace{6pt}  $^{a}${\em{Department of Physics \& Astronomy, University of California, Los Angeles, CA 90095 USA}};
$^{b}${\em{Department of Astronomy, California Institute of Technology, Pasadena, CA 91125 USA}}\\\vspace{6pt}\received{v3.0 released January 2010} }

\maketitle

\begin{abstract}
Gravitational lensing - the deflection of light rays by gravitating matter - has become a major tool in the armoury of the modern cosmologist. Proposed nearly a hundred years ago as a key feature of Einstein's theory of General Relativity, we trace the historical development since its verification at a solar eclipse in 1919. Einstein was apparently cautious about its practical utility and the subject lay dormant observationally for nearly 60 years. Nonetheless there has been rapid progress over the past twenty years. The technique allows astronomers to chart the distribution of dark matter on large and small scales thereby testing predictions of the standard cosmological model which assumes dark matter comprises a massive weakly-interacting particle. By measuring distances and tracing the growth of dark matter structure over cosmic time, gravitational lensing also holds great promise in determining whether the dark energy, postulated to explain the accelerated cosmic expansion, is a vacuum energy density or a failure of General Relativity on large scales. We illustrate the wide range of applications which harness the power of gravitational lensing,  from searches for the earliest galaxies magnified by massive clusters to those for extrasolar planets which temporarily brighten a background star. We summarise the future prospects with dedicated ground and space-based facilities designed to exploit this remarkable physical phenomenon.
\end{abstract}

\section{Introduction}

One of the most intriguing aspects of the propagation of light rays in
the cosmos is their deflection by massive objects. The phenomenon -
termed {\it gravitational lensing} was predicted almost exactly a century
ago by Albert Einstein as a feature of his theory of General
Relativity and has now become one of the most powerful tools of the
observational astronomer. We considered it appropriate, in this
celebration of the `Year of Light',  to provide a non-specialist review
of the progress that has been made utilising this phenomenon as well
as introducing ambitious plans by the international community for
future applications of gravitational lensing with upcoming facilities.

Einstein's prediction that light would be deflected by the Sun at the
time of a solar eclipse was first verified by the famous 1919
expedition of the British astronomers Arthur Eddington and Frank
Dyson. However, Einstein and Eddington were surprisingly skeptical of
the long term utility of gravitational lensing. A renaissance began
only in the 1970's when improved astronomical detectors and powerful
large telescopes became available. The exquisite image quality of
Hubble Space Telescope (HST), launched in 1990, provided a further
major boost in progress. 

Today gravitational lensing is being used to explore the distribution
and nature of the poorly-understood dark matter that provides the
dominant component of mass in the Universe. It is proposed that the
phenomenon can yield unique insight into the mysterious {\it dark energy}
- a property of space invoked to explain the accelerated expansion of
the cosmos discovered in the late 1990's. Foreground mass structures
such as nearby clusters of galaxies can also be used as natural
`telescopes' whereby distant objects are magnified as with a traditional
optical lens; this provides unique information about 
galaxies seen at early cosmic times; without gravitational lensing
such sources would be too faint to study. 

Here we will briefly discuss the fascinating history of this
phenomenon, explain in simple terms how gravitational lensing works,
review the progress in areas of contemporary interest, and conclude
with ambitious plans for future applications.  For interested readers
there are many excellent reviews of gravitational lensing
\cite[e.g.,][]{SKW06,Tre10,Bar10,K+N11}. A comprehensive commented
bibliography, including material suitable from high school to graduate
students is given by \cite{TMC12}.

\subsection{A remarkable history}

The earliest known mention of light being deflected by massive objects
is the first query in Newton's Opticks in 1704: 'Do not Bodies act
upon Light at a distance, and by their action bend its Rays; and is
not this action strongest at the least distance?'' Unfortunately, the
query does not distinguish between the action of gravity on a
corpuscle and more conventional optical phenomena. Henry Cavendish is
credited with the first (unpublished, 1784) calculation of the
deflection angle $\theta$ of a corpuscular light ray following a
hyperbolic trajectory and the origin of the (Newtonian) equation
$\theta = 2 GM / Rc^2$ where $M$ is the mass of the deflector and $R$
the radius at which the light ray arrives. Subsequently in 1804, Johann von Soldner
published a similar calculation deriving a deflection of 0.84 arcsec
for stars viewed close to the limb of the Sun. von Soldner
additionally discussed the practicality of verifying this prediction
but his work, as well as that of Cavendish, was largely forgotten as
the corpuscular theory of radiation was increasingly discredited in
favour of wave theories of light. Not only was there confusion as to
whether a deflection was expected for a light wave but the small
deflection was also considered unobservable.

In 1911, Einstein calculated a relativistic version of the solar
deflection and derived a similar result to that achieved by von
Soldner a hundred years earlier, 0.875 arcsec. However, the physical
principles behind the two calculations are quite different. In the
classical calculation, it is assumed that light can be accelerated and
decelerated like a normal mass particle, whereas in Einstein's
calculation the deflection is based on gravitational time dilation. In
1915, Einstein considered the additional deflection arising from the
curvature of space around the Sun in his newly-published General
Theory from which he derived $\theta = 4 GM / Rc^2$ and a solar
deflection of 1.75 arcsec. Beginning in 1912, Einstein sought
observers who could verify his predicted deflection. The observational
race to prove or disprove Einstein's theory is a fascinating story
well-documented in several recent books
(e.g. \cite{Col99}\cite{Cre06}\cite{Sta07}\cite{Gat09}).

The Astronomer Royal, Frank Dyson, first proposed the May 29th 1919
eclipse expedition noting that the Sun would be in the rich field of
the Hyades star cluster. Arthur Eddington had played a key role in
promoting Einstein's theory and took the lead in the
organization. Eddington and his assistant Cottingham visited the
island of Pr\'incipe off the coast of West Africa (now part of the
democratic republic of Sao Tom\'e and Pr\'incipe); another team
(Crommelin and Davidson) visited Sobral, Brazil. The results,
confirming the full deflection, were presented in November 1919
(\cite{Dys20}). Although some have argued that Eddington was blinded
by his enthusiasm for Einstein's theory and biased in his analysis by
discarding discrepant data \cite{Wal02}, a recent re-analysis by
Kennefick \cite{Ken07} shows this was not the case. The rejected
Sobral astrograph plates were out of focus as a result of the rapid
change in temperature during totality making it difficult to establish
a proper plate scale.  In 1979 the Sobral plates were more accurately
re-measured yielding a deflection of 1.55 $\pm$ 0.32 arcsec
\cite{Har79}.

Eddington and Einstein were curiously reticent about possible
applications of gravitational lensing. Chwolson \cite{Chw24}
illustrated how lensing can produce multiple images of a distant
source -- a phenomenon now termed strong lensing (\S2) but, as its
occurrence depends on the precise alignment of a source and deflector,
it was reasonable to conclude the probability of observing such
phenomena would be very small. As a good illustration of thinking at
the time, Einstein, urged by Mandl, discussed what Paczynsky later
called microlensing -- the temporary brightening of a star due to the
magnification induced by a foreground object that crosses the line of
sight to the observer (\S4). In this rare post-1919 article about
lensing by its predictor \cite{Ein36}, he states ``of course there is
no hope of observing this phenomenon.''

Fritz Zwicky emerges as the worthy prophet by arguing in 1937 that
galaxies and galaxy clusters would be far more useful lenses given
their greater mass and cross-section to background objects and, with
great vision, foresaw many of the applications we review here
\cite{Zwi37}. In the 1960s, Barnothy \& Barnothy \cite{Bar68} became
tireless advocates of Zwicky's position. The mathematics of
multiply-imaged geometries was further developed independently by
Klimov \cite{Kli63}, Liebes \cite{Lie64} and Refsdal. Refsdal
\cite{Ref64} demonstrated that if a background lensed source such as a
quasar or supernova is variable in its light output, an absolute
distance scale can be determined by measuring the time delay in the
arrival of light observed in its multiple images; this offers a
geometric route to measuring the rate of expansion of the Universe.

\begin{figure*}
\centerline{
\psfig{file=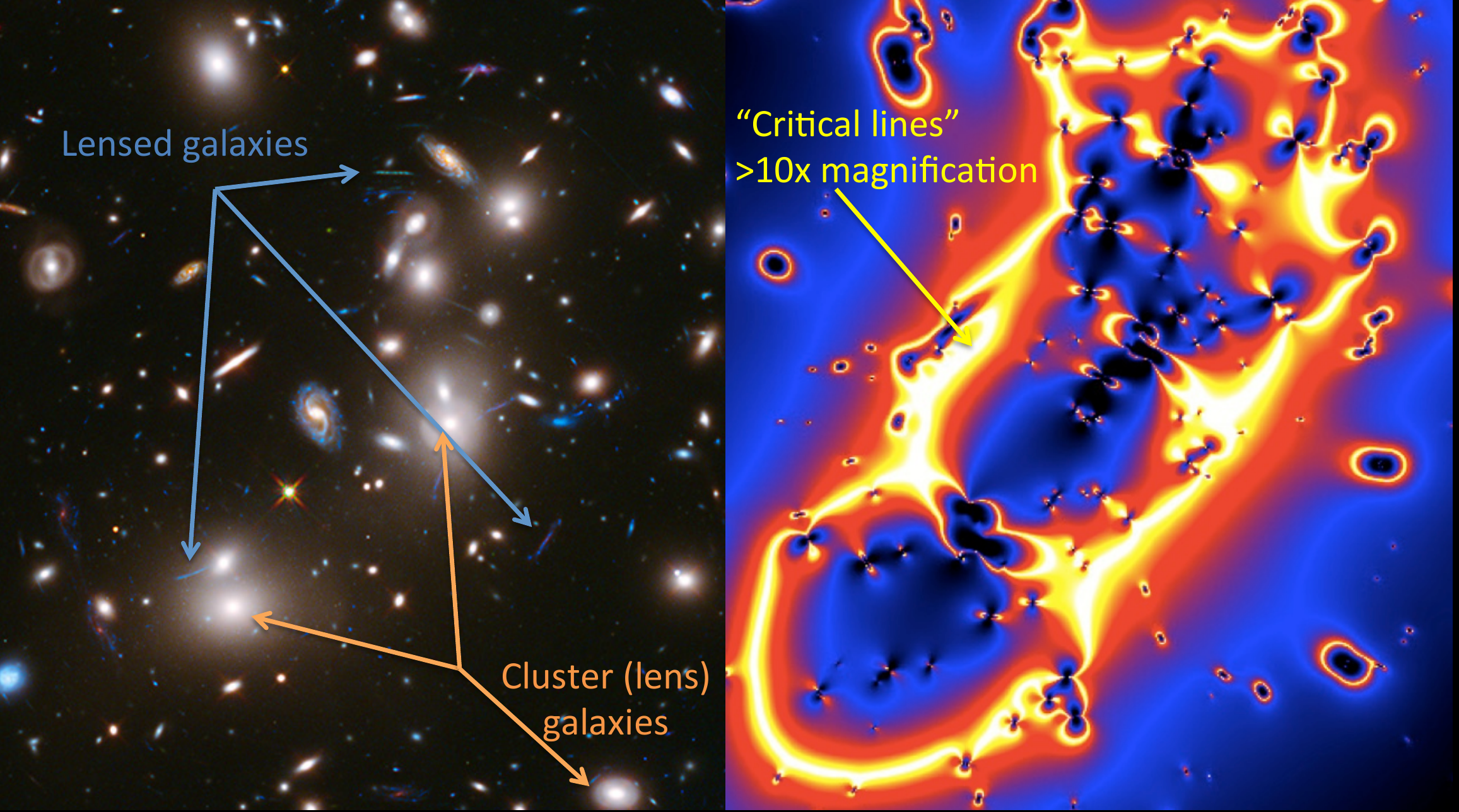,width=0.9\textwidth}}
\caption{A striking example of gravitational lensing by a cluster of galaxies. Left: Hubble Space Telescope 
false colour image of Abell 2744 revealing many luminous members (white/yellow color)
but also numerous background galaxies (typically blue) stretched and
distorted by the gravitating mass in the cluster which is dominated by
dark matter. The image is taken as part of the Frontier Fields
programme .  Right: Magnification map inferred via a gravitational
lens model for the cluster
\cite{Jau++14b}. The cluster acts as a natural telescope, so that a
distant background source would appear magnified by a factor of
$\sim10$ or more if it were to appear near the so-called critical
lines (the regions surrounding the cluster centre colored in
yellow/white,where magnification is very high, formally infinite for a
point source). Sources in most of the solid angle behind the cluster
are magnified by a factor of a few.  Credits: NASA/ESA and
\cite{Jau++14b}. \label{fig:a2744}}
\end{figure*}

Why did it take so long before further observational progress was made
in gravitational lensing? Firstly, gravitational lensing is a relative
rare phenomenon in the celestial sky requiring fortuitous alignment of
foreground and distant sources. Secondly, as in conventional optics,
the background source must be substantially more distant than the
lens. Until the 1960s, very few truly cosmologically-distant sources
(i.e. at large redshift\footnote{`Redshift' is defined as the factor
by which the wavelength of the light of a source is stretched by the
cosmic expansion and hence is a valuable measure of its distance, and
`look-back' time.}) were known. Only as quasar surveys yielded many distant
sources in the 1970s did it finally become likely one would be found
behind a foreground galaxy. The first example, SBS 0957+561 A/B, was
verified spectroscopically by Walsh, Carswell \& Weymann in 1979
\cite{Wal79}. They discovered two images of the same distant (redshift
$z$=1.413) quasar gravitationally-lensed by a foreground galaxy with
redshift $z$=0.355. Thirdly, surface brightness is conserved in the
lensing process (as in conventional optics). However, as surface
brightness dims with increased redshift $z$ as $(1 + z)^{-4}$ due to
relativistic effects associated with the expansion of the Universe,
many lensed images viewed through foreground galaxy clusters lay
undiscovered until the 1980s when efficient digital cameras became
commonplace on large ground-based telescopes. The increased
sensitivity led to the discovery in the mid 1980s of many `giant arcs'
- distorted images of background galaxies. For a few years there was
some speculation as to the origin of these strange
features. Eventually, Soucail and colleagues \cite{Sou++88} confirmed,
with a spectrum, that a giant arc in the rich cluster Abell 370 at a
redshift $z$=0.37 is the distorted image of a single background galaxy
at redshift $z$=0.724. The improved angular resolution of the Hubble
Space Telescope (HST), launched in 1990, was later critical in
recognizing numerous distorted images of faint sources. HST features
very prominently in science programmes exploiting gravitational
lensing, for example via its role in conducting deep imaging surveys
of foreground clusters for highly-magnified distant objects
(Figure~\ref{fig:a2744}).

\subsection{Gravitational lensing and Fermat's principle}

\begin{figure*}
\centerline{\psfig{file=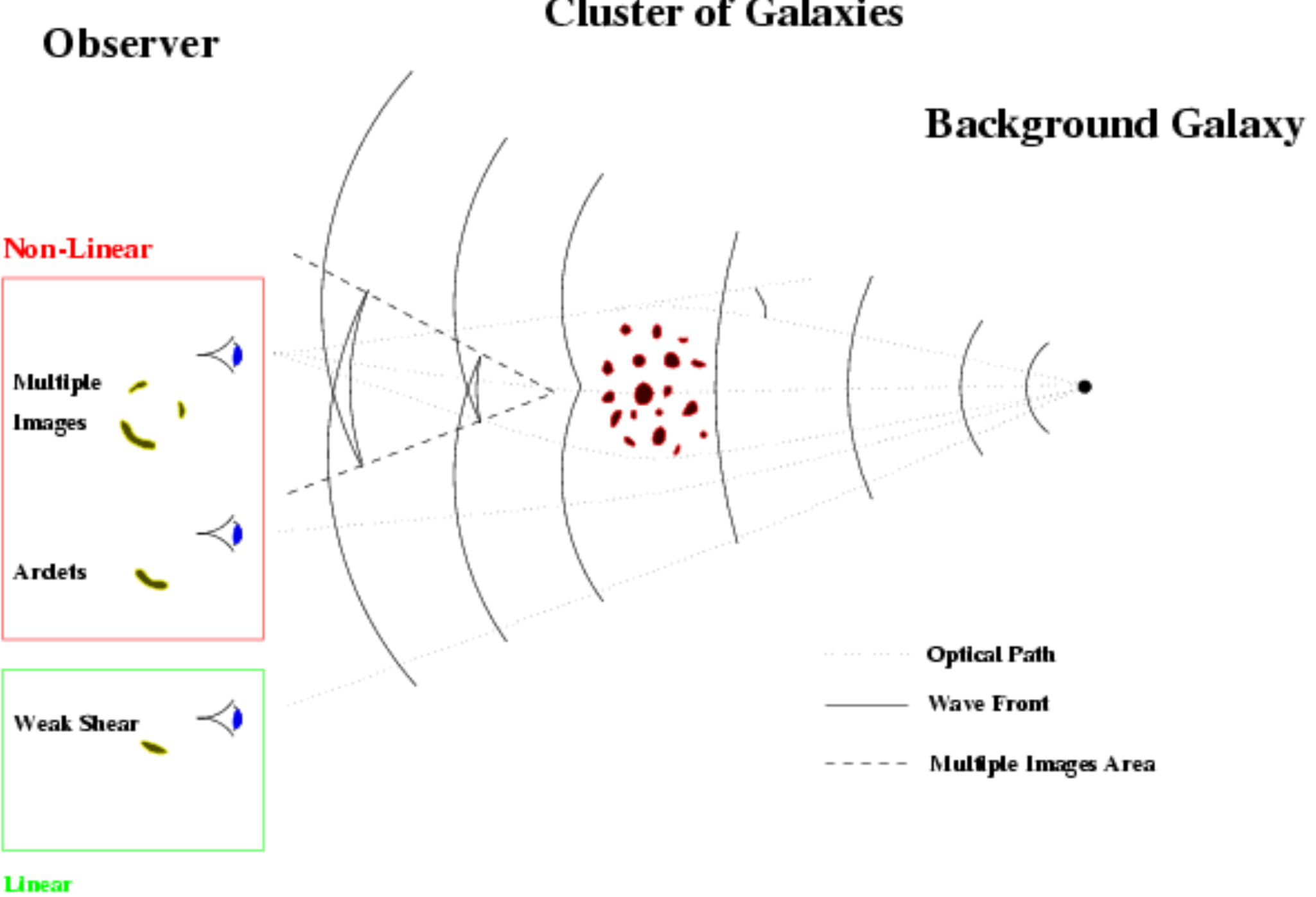, width=0.8\textwidth}}
\caption{How gravitational lensing works: a foreground cluster of galaxies contains copious amounts of mass (mostly dark matter) and acts as a gravitational lens, distorting and magnifying the light of a background galaxy. The resulting image seen by the observer depends on the relative distances between the observer, lens and source, the concentration of mass (dark and visible) in the lens and the degree of alignment of the observer, lens and source. When the alignment and mass concentration is sufficient to create multiple-images of the background source, the phenomenon is called {\it strong lensing}. In this case the magnification and distortion can be very significant. When the alignment is not so fortuitous, only a modest stretching of a single image may occur ({
\it arclet} in the case shown). Quite generally dark matter in the cosmos induces a small distortion in the shapes of all background galaxies - a phenomenon called 
{\it weak lensing} or {\it cosmic shear}. \label{fig:sketch}}
\end{figure*}

\begin{figure*}
\centerline{\psfig{file=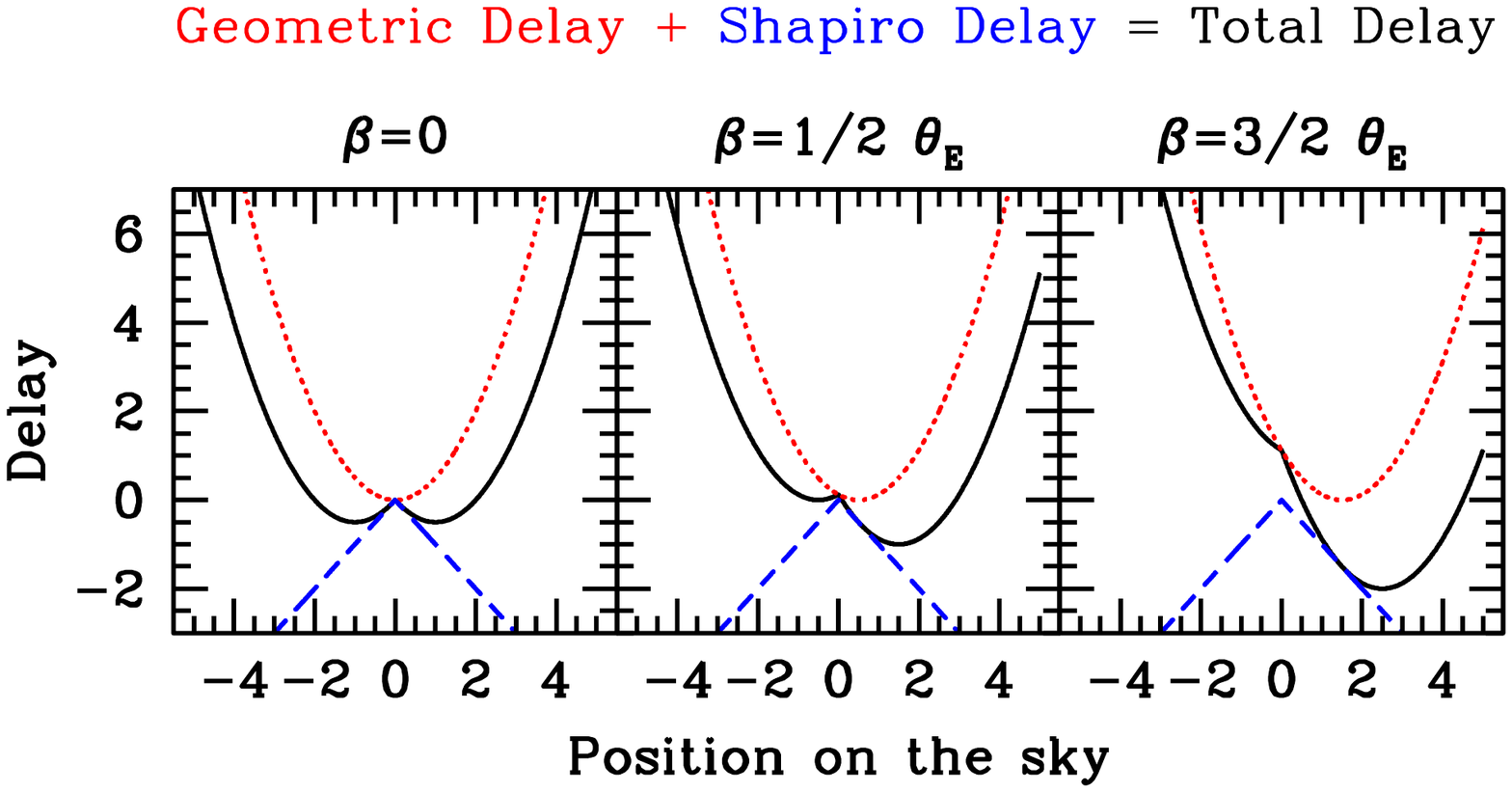, width=0.8\textwidth}}
\caption{Illustration of gravitational lensing in terms of Fermat's principle. For a given source position ($\beta$), the time delay surface (black solid lines) is given by the sum of the geometric delay (red dotted lines) and the Shapiro delay (blue dashed lines) as a function of position in the image plane ($\theta$, in units of the Einstein Radius $\theta_{\rm E}$, typically one arcsecond for galaxy-scale lenses). Images then form at the extrema of the time delay surface. The three panels show a section of the time delay surface in three different regimes of circularly symmetric deflector. Left panel: the source is perfectly aligned with the deflector ($\beta=0$); the time delay has a local maximum at the center and two minima at the same height. This configuration gives rise to a perfect Einstein Ring, with an infinitely demagnified image in the center (see centre panel of Figure~\ref{fig:Sonnenfeld} for example). Middle panel: the source is now offset to one side by half an Einstein Radius; the image forming on the outer minimum arrives first, then the central image corresponding to the local maximum, and last the image corresponding to the inner minimum. This configuration gives rise to the classic double configuration, with two bright images and an infinitely demagnified central image. Right panel: the source is now offset by more than the Einstein Radius; in this case there is only a minimum and thus only one image, i.e. no strong lensing.
\label{fig:delays}}
\end{figure*}

Figure~\ref{fig:sketch} provides a useful qualitative overview of
the lensing phenomenon. When the lens, for example a foreground galaxy
or cluster of galaxies, is dense enough and well-aligned with a
background source, multiple images of that source can be
produced. More generally, when the intervening mass is less
concentrated and/or poorly-aligned with the background source, only a
small shape distortion occurs.

Quantitively, gravitational lensing can be described in a similar
manner to that of standard optics. The formulation of gravitational
optics in terms of a generalized version of Fermat's principle
provides a very intuitive way to understand the underlying physics and
the basic phenomenology. We give here a brief description of
gravitational optics referring to \cite{Tre10} for a summary with
equations and to \cite{B+N92} for a more thorough discussion.

The mathematic formalism of gravitational lensing can be conveniently
described as a transformation between the positions and shapes of
background sources as they would be observed in the absence of a
deflector (the so-called {\it source plane}) and that received by
observer (the so-called {\it image plane}). The transformation is
achromatic and preserves surface brightness \cite{Bar10}.

The transformation from source to image plane is given by Fermat's
principle. As in conventional optics, photons seem to ``choose''
special paths from the source to the observer, i.e. images form at the
extrema of the arrival time surface (usually they pick the shorter
time, i.e. minima, but also maxima or saddle points are allowed).
However, whereas in standard optics the light travel time depends on
the speed of light in the material as well as on the path length, in
gravitational optics the role of the refractive material is played by
the gravitational potential of the deflector via the so-called Shapiro delay
\cite{Sha64}, i.e. the delay measured by the observer for a light
ray passing through a deep gravitational potential caused by general
relativistic time dilation. The competition between the Shapiro delay, which
increases with the gravitational potential, and the length of the
light path gives rise to the variety of observed phenomena.  The
description of gravitational lensing in terms of Fermat's principle is
illustrated in Figure~\ref{fig:delays}.

Under rare circumstances, if the Shapiro delay is strong enough,
multiple images can appear to the observer, giving rise to the
phenomenon of {\it strong lensing}. In this case the time delay
between the arrival of light to the various images encodes information
about the absolute path lengths traversed and hence the size of the
Universe as function of cosmic time. As we describe in
Section~\ref{ssec:de}, this provides an opportunity for a direct
measurement of various cosmological parameters.  Conversely, if the
Shapiro delay is not strong enough to counterbalance the geometric
delay, only a single distorted image of the source appears to the
observer. This phenomenon is called {\it weak lensing} and is a
powerful tool to trace the distribution of dark matter outside of the
confines of massive structures like cluster of galaxies
(Section~\ref{sec:weak}).

\section{Strong Lensing}

Strong lensing is perhaps gravitational lensing's most visually
impressive feature. A rich cluster of galaxies such as that in Figure
1 produces striking distorted images of background galaxies which
appear to swirl around the cluster core. Importantly the phenomenon is
governed by the {\it total} mass in the cluster whether it is visible,
as in the member galaxies, or dark. As we know the bulk of the
gravitating matter in the Universe is in fact invisible, lensing
offers us a remarkably powerful tool to study both the distribution
and nature of dark matter.

\subsection{Lensing anomalies and the nature of dark matter}

The standard picture of dark matter is that it is comprised of a
massive weakly-interacting or `cold' particle. We know it cannot be
baryonic (i.e. quarks) in form as this would violate measured
abundances of the light elements synthesized in the Big Bang
\cite{Ste06}, and the observed power spectrum of cosmic microwave
background \cite{Hin++13}. Although physicists can attempt, with
difficulty, to capture this weakly-interacting particle and constrain
its mass and properties, astronomers have a unique ability to observe
the dark matter using gravitational lensing.

A very robust prediction of the standard cold dark matter (CDM)
cosmological model is that dark matter congregates in large `halos'
within which are numerous satellites or `subhalos'. The abundance of
these subhalos should increase rapidly with decreasing mass (dN/dM$_{
\rm sub}\propto M_{\rm sub}^{-1.9}$) \cite{Spr++08}. 
This behaviour stems directly from the `coldness' (i.e. low thermal
speed) of dark matter in the standard model. Alternate cosmological
models with less massive (or `warmer' ) dark matter particles
\cite[e.g. keV scale sterile neutrino][]{Kus09} predict a lower mass 
cutoff to the distribution of subhalos
\cite{Nie++13,Lov++13,Ken++14}.

The luminous satellites surrounding our Milky Way and external
galaxies do not appear to be nearly as abundant as the predicted
distribution of subhalos in CDM, a discrepancy dubbed the `missing
satellite problem'\cite{Moo++99,Kly++99}(see also \cite{BBK11} for an
associated problem).  The most favoured solution is that the lower
mass subhalos cannot retain their hydrogen gas and are thus unable to
form stars or be seen. This implies that there should be thousands of
dark subhalos orbiting our own Milky Way.

\begin{figure*}
\centerline{
\psfig{file=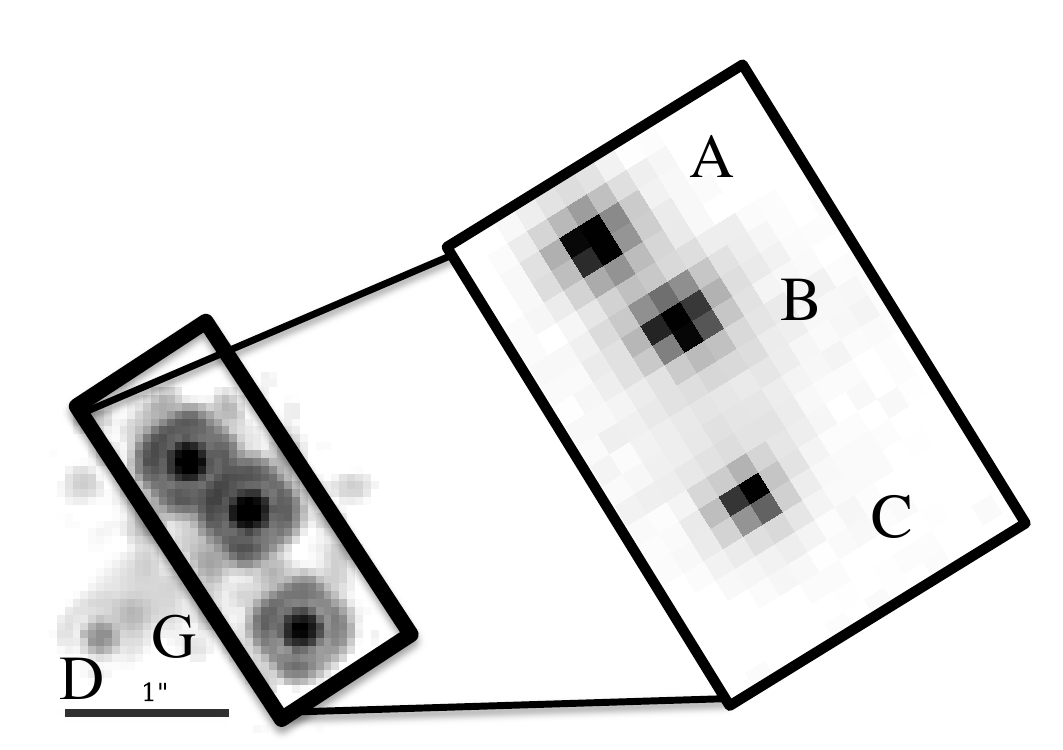, height=6cm}
\psfig{file=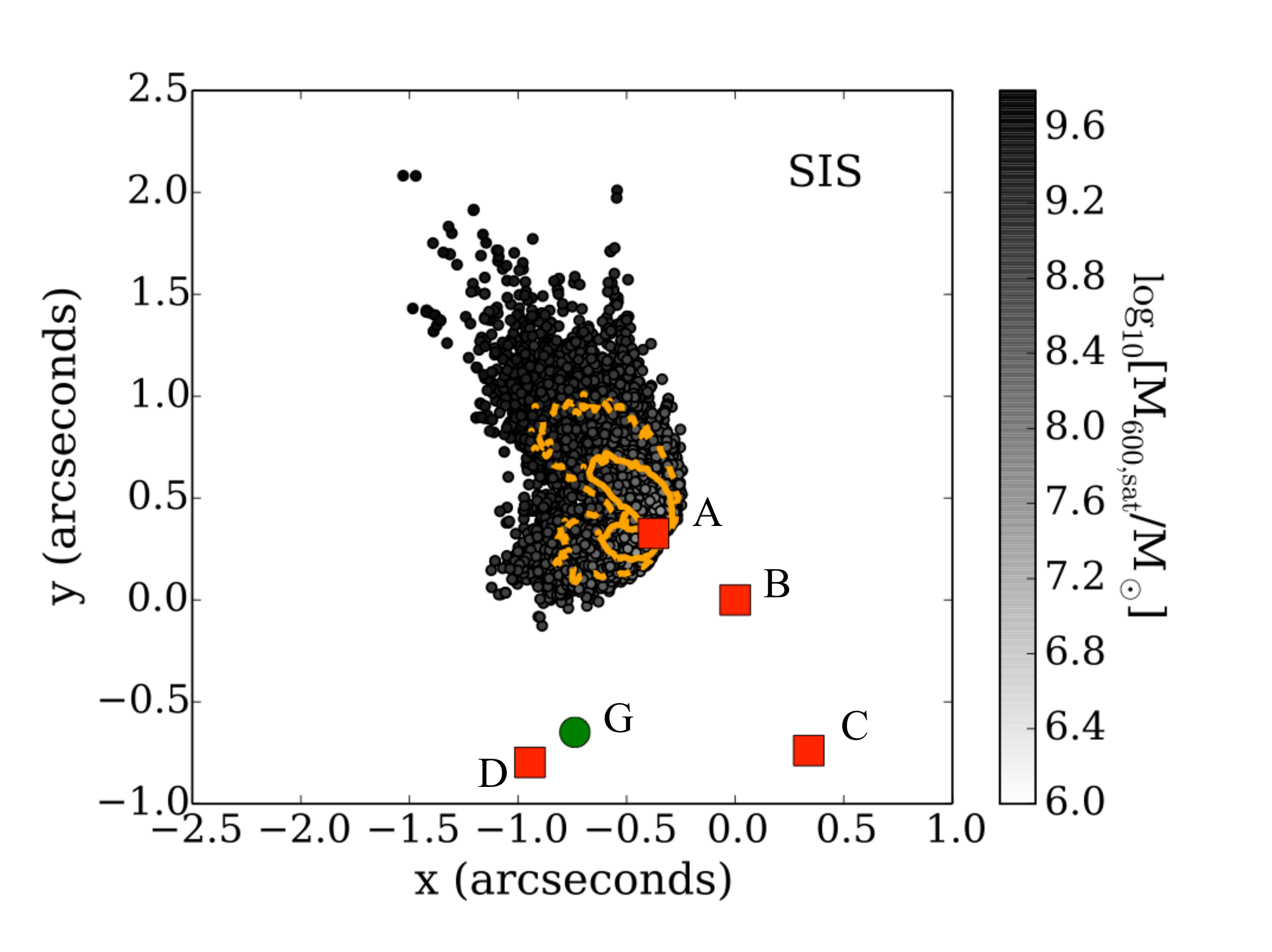, height=6cm}}
\caption{Detecting dark matter substructures to test the standard cold dark matter model. The left panels show four multiple images (A,B,C and D) of the gravitationally-lensed quasar B1422+231 obtained with the Hubble Space Telescope (leftmost panel) and with the imaging spectrograph OSIRIS behind the adaptive optics system on the W.~M.~Keck-I Telescope (zoomed in panel). The relative positions and narrow line spectroscopic fluxes of these images measured with OSIRIS contain clues as to the distribution of dark matter in the foreground lens (G). Accurate modelling indicates the requirement for a sub-halo, in addition to the primary one, whose relative position and mass are constrained as shown with the distribution of black points and orange contours in the right panel. Such positional and flux anomalies can be used to trace dark matter sub-halos with masses of $\simeq10^8$ solar masses, providing a critical test of the standard model (after \cite{Nie++14}). 
\label{fig:Nierenberg}}
\end{figure*}

Given the uncertainties in understanding star formation in low-mass
galaxies, it is clear that only a direct census of these subhalos by
mass can tell us conclusively whether these satellites do not exist or
whether they are simply dark. This is a remarkably clean measurement
in principle: if the dark subhalos do not exist, the standard cold
dark matter model would be ruled out. Gravitational lensing provides a
unique opportunity to perform this measurement, by means of 
so-called {\it strong lensing anomalies}. The presence of dark unseen
satellites can be detected as small scale perturbations in the
gravitational potential of a massive galaxy
\cite{M+S98,M+M01,K+M09}. These perturbations with respect to an otherwise 
smooth mass distribution change the arrival time, apparent position,
and observed flux of the lensed sources (hence they are named
time-delay, astrometric, and flux ratio anomalies, respectively). An
illustration of one of the methods is shown in
Figure~\ref{fig:Nierenberg}.

The results so far indicate that the abundance of dark subhalos is
consistent with the expectations of CDM models, albeit with large
uncertainties \cite{D+K02,Xu++09,Veg++14,Xu++14}. This is an area
where great progress will be possible in the next decade, by studying
large numbers of strong lenses at high angular resolution.

\subsection{Time delays as a probe of dark energy}
\label{ssec:de}

If the mystery of dark matter was not enough of a problem for the
cosmologist, the discovery of the accelerated expansion of the
Universe in 1999 from the study of distant supernovae
\cite{Per++99,Rie++98} raises a new conundrum. Over 70\% of the energy
density in the Universe is contained in the so-called {\it dark energy} - a
label used to cover our ignorance of one of the most basic features of
the Universe. Contemporary thinking suggests dark energy may be a
natural property of empty space, a {\it vacuum energy density}, possibly
the cosmological constant initially invoked by Einstein to retain
a static Universe. A key question in considering the nature of dark
energy is whether it is a constant property in cosmic time or whether
it evolves. This is central to understanding the fate of the Universe.

A powerful method to determine the nature of dark energy is to measure
the time evolution of cosmic distances. In the same way as the
strength of the Earth's gravitational field can be inferred from the
trajectory of a football, the evolution of physical scales in the
universe provides information about its total energy density

Distance measurements are critical in cosmology. The supernovae
observations that led to the surprising discovery of an accelerating
Universe measured the relative distances between supernovae at
different redshifts. Likewise, one of the fundamental quantities
measured by cosmic microwave background satellites such as Planck
\cite{Pla++13} and WMAP \cite{Hin++13} is the distance to the last 
scattering surface of the cosmic microwave background at redshifts
$\sim1100$ (when the universe was 370,000 years old).

Strong gravitational lensing of a variable source provides a very
elegant one step measurement of absolute distances in the Universe.
The difference in arrival time induced by a lens is given by $\Delta t
\propto D_{\Delta t}{\delta\phi}$. Here $D_{\Delta t}$ is the so-called time 
delay distance and encompasses all of the cosmological dependence, and
$\delta\phi$ describes the geometry of the system and the
gravitational potential of the main deflector. By measuring a time
delay and determining a mass model for the main deflector, one obtains
the time-delay distance $D_{\Delta t}$ and, thus, a determination of
the cosmological parameters \cite{Ref64,Sch++97}.

\begin{figure*}
\centerline{\psfig{file=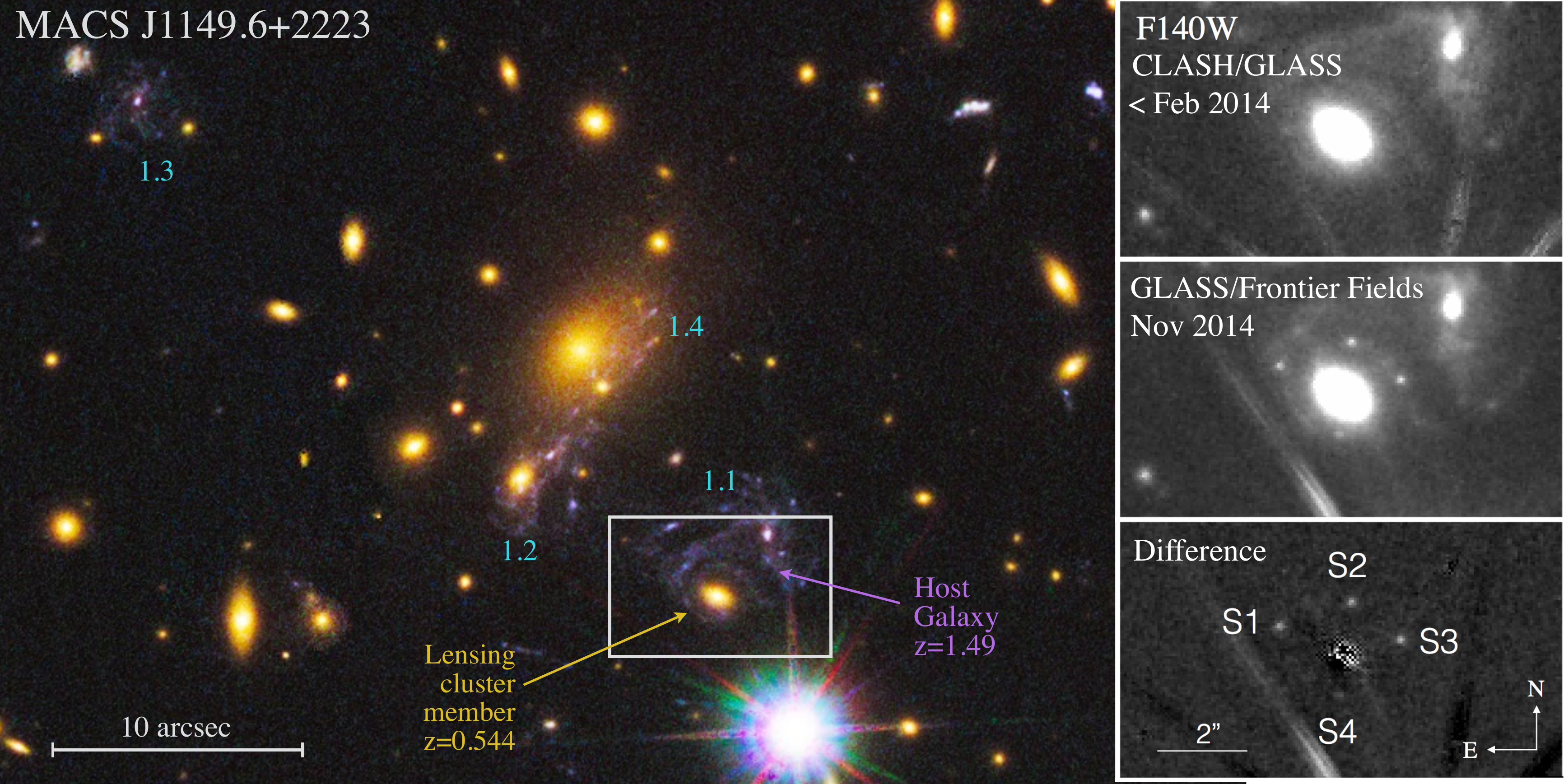, width=0.99\textwidth}}
\caption{Discovery of the multiply-imaged supernova 'Refsdal'. The left panel shows an image of the cluster of galaxies MACS J1149.6+2223 taken prior to the explosion of the supernova. The host galaxy at $z=1.49$ is multiply imaged by the cluster, forming images 1.1, 1.2, 1.3, 1.4. A portion of the host galaxy is further multiply imaged by a galaxy in the lensing cluster at $z=0.544$. The right panels zoom in on the  multiply imaged supernova, (top) image prior to the supernova explosion from the CLASH program (PI: Postman), (middle) discovery image from the GLASS program (PI: Treu), (bottom) difference image revealing four images of the supernova in an `Einstein Cross' configuration. Left panel image credit: NASA, ESA, W. Zheng (JHU), M. Postman (STScI), and the CLASH Team. The right panel image is taken from \cite{Kel14b}. Montage and labels courtesy of S.A.~Rodney.
\label{fig:Refsdal}}
\end{figure*}

\begin{figure}
\hbox{
\psfig{file=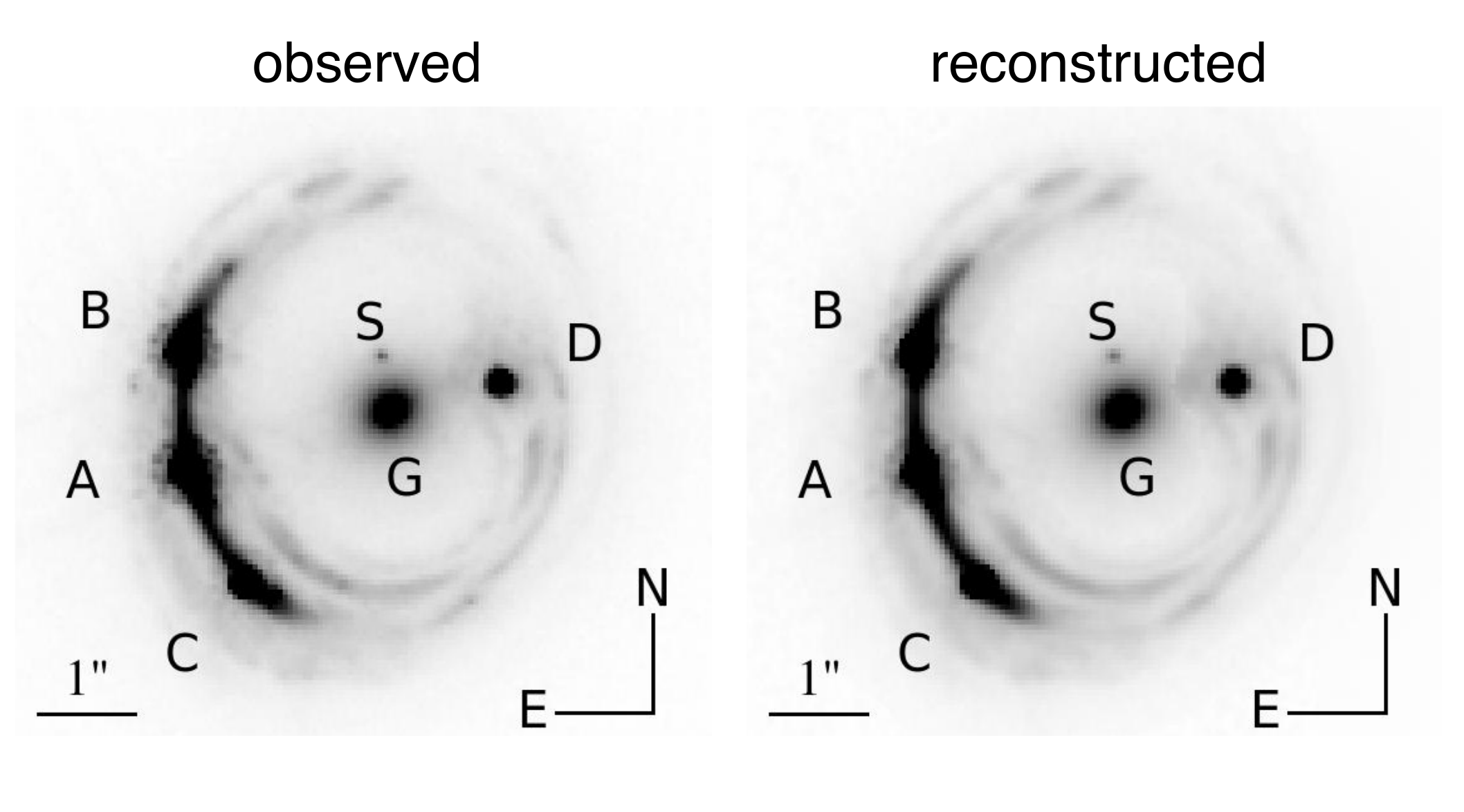, width=0.6\textwidth} 
\psfig{file=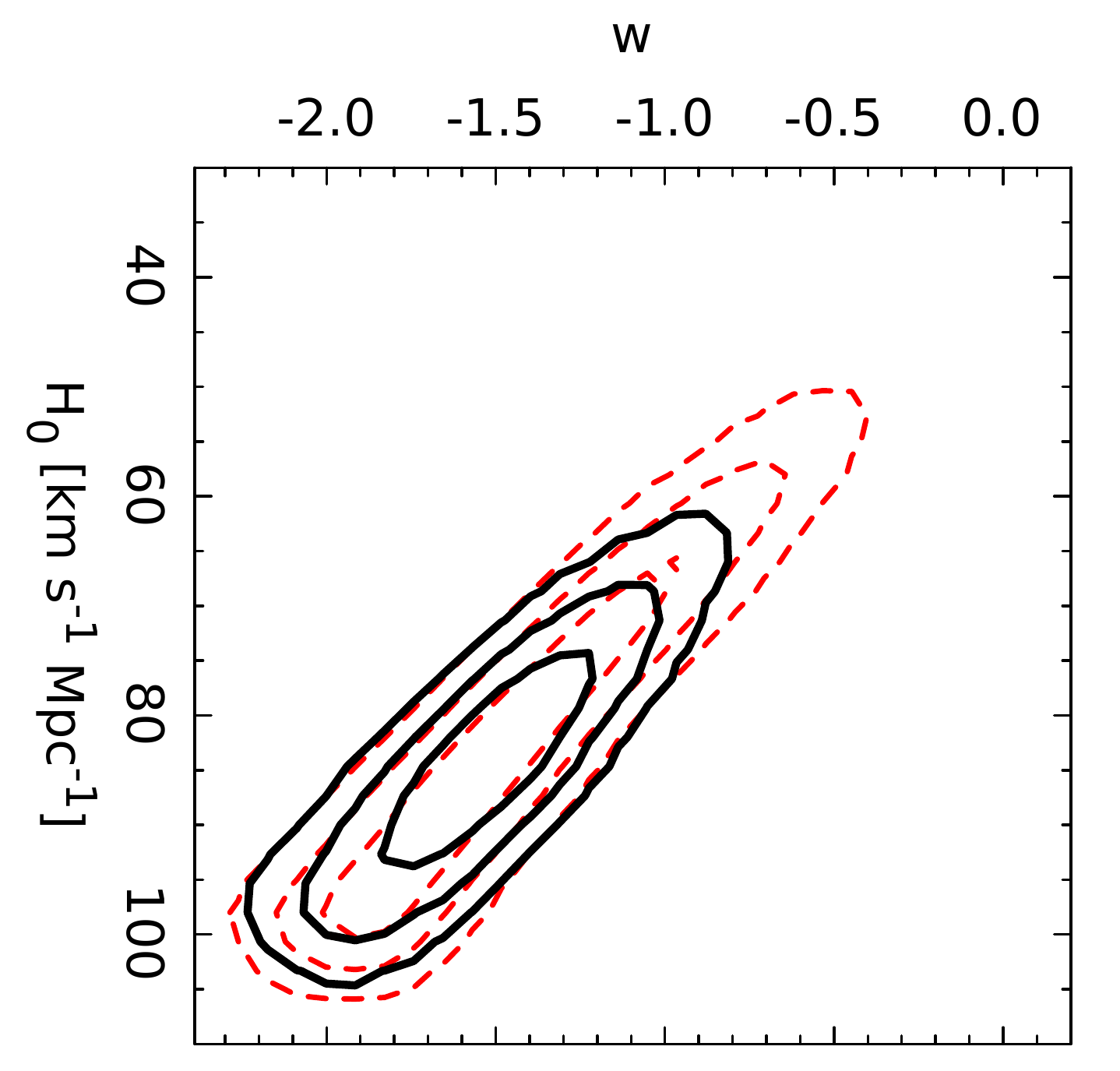, width=0.4\textwidth} 
}
\caption{Constraining the nature of dark energy using gravitational time delays. The leftmost panel shows a Hubble Space Telescope image of the multiply-imaged variable quasar RXJ1131-1231 gravitationally-lensed by a foreground galaxy G and its associated satellite S.  To its right is a reconstructed image based on a mass model for the lens. Accurate monitoring of the three images A, B and C provide an absolute measure of the different path lengths for a light ray through the lens and hence constraints on the present expansion rate of the Universe (Hubble's constant $H_o$) and the equation of state parameter ($w$; i.e. the ratio between pressure and energy density, $w=0$ for cold dark matter, $w=1/3$ for photons, $w=-1$ for a cosmological constant) of dark energy as shown in the rightmost panel (after \cite{Suy++14}). The dashed contours show the 68\%, 95\%, and 99
\% posterior probability density contours based on the cosmic microwave data alone, and the solid contours show the improved precision with the inclusion of the time-delay distance measured from RXJ1131-1231.}
\label{fig:Suyu}
\end{figure}

Refsdal's idea \cite{Ref64} is fifty years old, but has only
very recently been realized as a practical proposition. His original
suggestion involved observing a multiply-imaged supernova, a rare
phenomenon. In an exciting development, the first example was
observed in 2014 \cite[][(discovery images of the long awaited
supernova aptly named 'Refsdal' are shown in
Figure~\ref{fig:Refsdal}]{Kel14a,Kel14b}. A more productive
endeavour, given the rarity of supernovae, has been the application 
of Refsdal's method to lensed variable quasars. Earlier efforts were first
stymied by the shortage of these sources,
and later by the logistical challenges associated with the necessary
long-term monitoring of them to measure accurate time delays. However, in the
past few years, dedicated monitoring efforts
\cite{Sch++97,Bar97,Bur++02,Fas++02,Par++09,Tew++13,Rat++13} and
advances in time delay measurements \cite{Pel++96,TCM13} and lens
modeling \cite{KKM01,T+K02b,W+D03,Koo++03,Suy++10} have led to
substantial progress. Recent work has shown that a single lens is
sufficient to measure absolute distances to 6\% precision
\cite{Suy++14} and thus determine whether dark energy is the
cosmological constant or a more exotic phenomenon
(Figure~\ref{fig:Suyu}). In the next decade with many planned wide
field surveys and dedicated efforts, thousands of lensed quasars will
be discovered and studied, yielding some of the most stringent
constraints on the properties of dark energy
\cite{Lin11,Tre++13}.

\subsection{Stellar and dark matter in massive galaxies}

\begin{figure}
\centerline{
\psfig{file=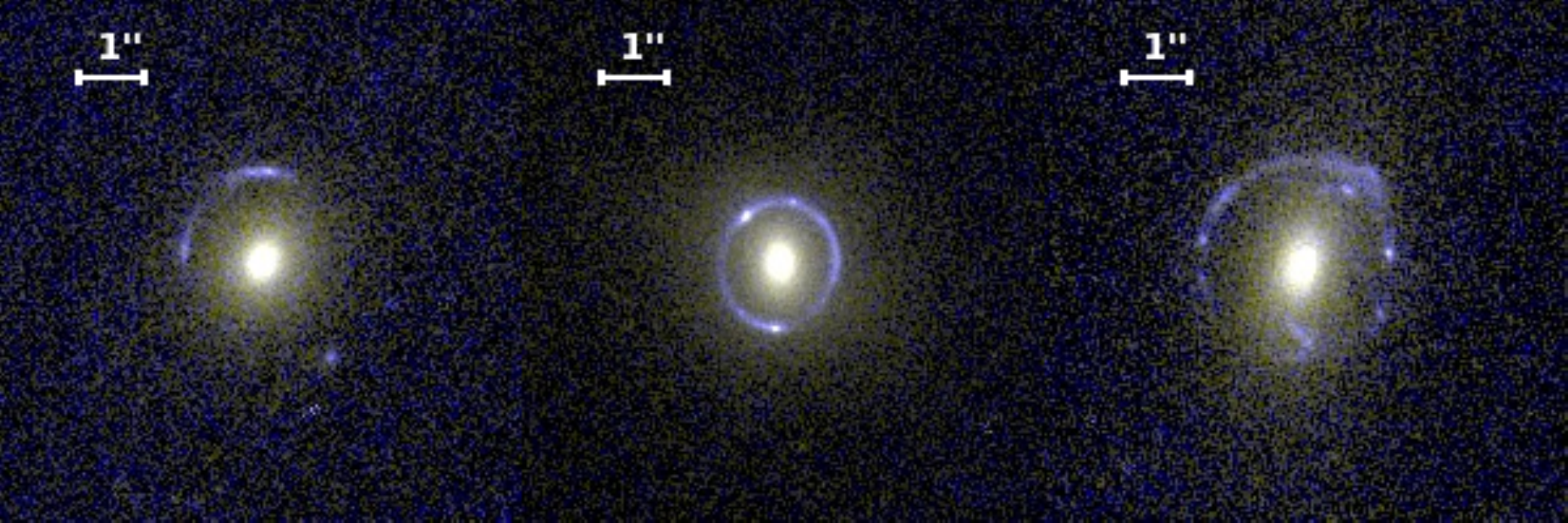, width=0.8\textwidth}}
\caption{Examples of strong gravitational lensing by a galaxy. The background sources (blue) are lensed by the foreground massive galaxy (yellow) into an almost perfect `Einstein ring'. The radius of the ring gives the total mass enclosed with a precision of just a few percent. Combining this geometrical measure with the kinematics of the stars in the foreground galaxy, deduced spectroscopically, provides key data on the radial distribution of matter in the lens as well as the nature of the stellar population (after \cite{Son++13a}). 
\label{fig:Sonnenfeld}}
\end{figure}

The idea that all galaxies are surrounded by halos of dark matter became
commonplace by the early 1980s. But how can we quantify the
distribution of dark matter around galaxies and verify its role in
galaxy formation given it is invisible? Elliptical galaxies are
compact and dense and thus serve as excellent gravitational lenses
\cite{SKW06}. Using spectroscopic data from the Sloan Digital Sky Survey,
the Sloan Lens Advanced Camera for Surveys (SLACS) team has so far
isolated over 100 elliptical galaxies that strongly lens background
blue star forming galaxies at $z=0.5-1$
\cite{Bol++08,Shu++14}. Since the redshifts of both the lens and background
source are known, the lensing geometry, revealed by Hubble Space
Telescope images (see examples of lenses in
Figure~\ref{fig:Sonnenfeld} taken from the SL2S survey
\cite{Gav++12}), defines the total mass interior to the so-called
Einstein radius irrespective of whether that material is
shining. Together with a dynamically-based mass on a smaller physical
scale derived from the dispersion of stellar velocities in the lensing
galaxy itself, the total mass density in the lens as a function of
radial distance within the galaxy, $\rho(r)$, can be determined.

Across a wide range in cosmic time and lens mass, the total mass
distribution is remarkably uniform following an isothermal
distribution, $\rho(r) \propto r^{-2}$ \cite{Son++13b,Bol++12}. This
distribution is spatially more extended than that of the visible
baryons demonstrating clearly the existence of dark matter
\cite{Koo++09}. These important results confirm that the early
formation of massive dark matter halos played the key role in
encouraging a rapid formation of the cores of massive galaxies.

One surprising result from this line of inquiry is that the ratio
between stellar mass and luminosity is much higher in massive
elliptical galaxies than in the Milky Way
\cite{Tre++10,Aug++10b,Cap++12,Son++12,Sche++14}. This finding, in
combination with detailed studies of the spectra of stars
\cite{CvD12}, suggest that mode of star formation in the most massive
galaxies is different than in typical galaxies, and yields much many
more low mass stars, down to the hydrogen burning limit
\cite{Bar++13}.

\subsection{The inner regions of dark matter halos}

A further area of tension between the standard cosmological model and
observations is the inner regions of galaxies and clusters of
galaxies. Dark matter only simulations predict that the density of
dark matter should be `cuspy'', i.e. increase at small radii as
$\rho_{\rm DM}\propto r^{-1}$ \cite{NFW97}, whereas observations of
many galaxies indicate a variety of density profiles including
so-called `cores'  where the dark matter density is constant within a
certain radius \cite{Don++09}. The origin of this discrepancy is not
well-understood because of the poorly understood physics of gas, stars
and black holes at the centres of galaxies. On the one hand, gas has
the ability to cool and sink toward the centre of the dark matter
potential wells; this makes the observational discrepancy even more of
a problem. On the other hand, the energy released by explosive events
such as supernovae and accretion onto black holes may potentially
mitigate the problem by transforming the steep cusps into cores
\cite{Oh++11,P+G12}.

\begin{figure*}
\hbox{
\psfig{file=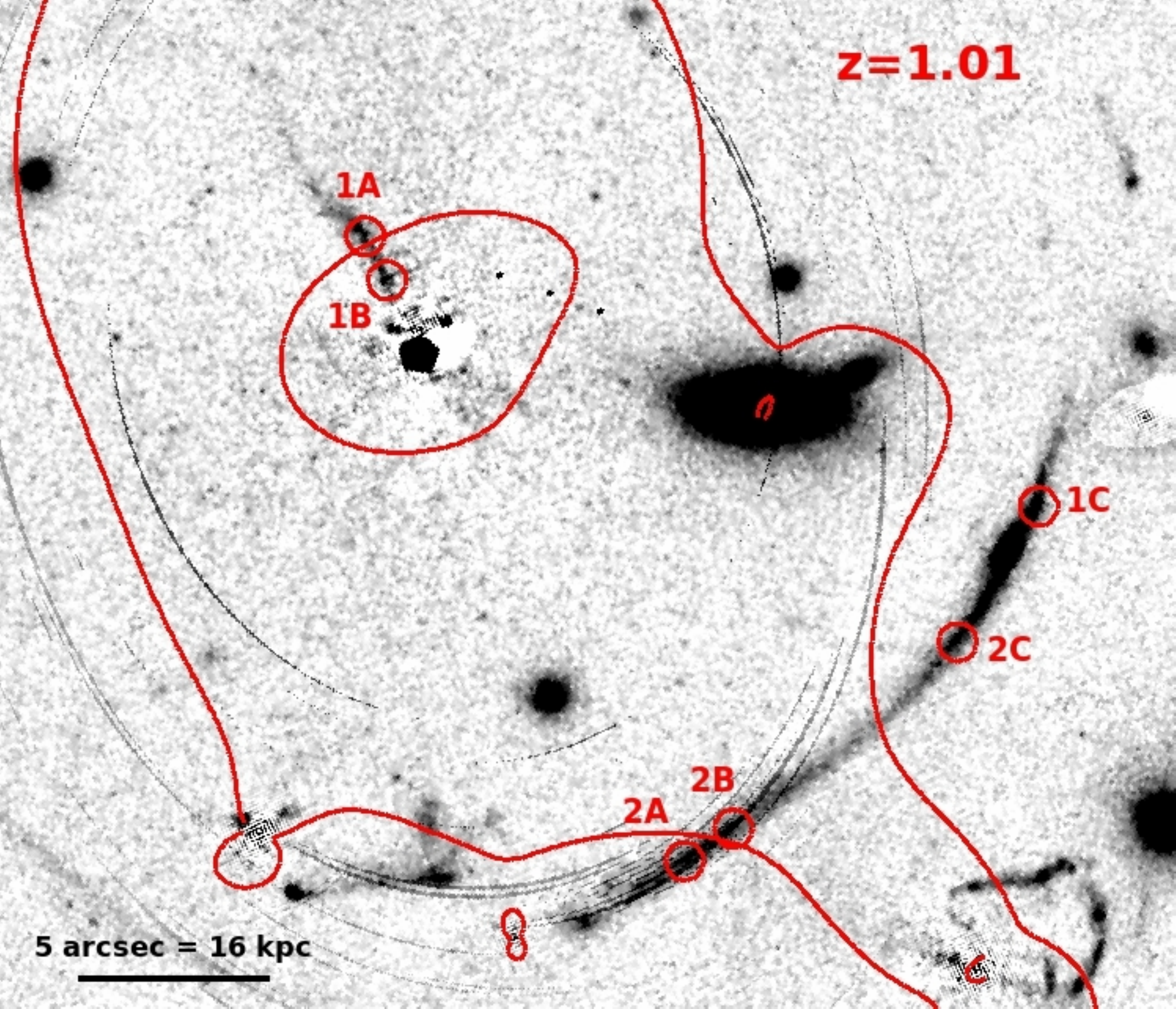, width=0.4\textwidth}
\psfig{file=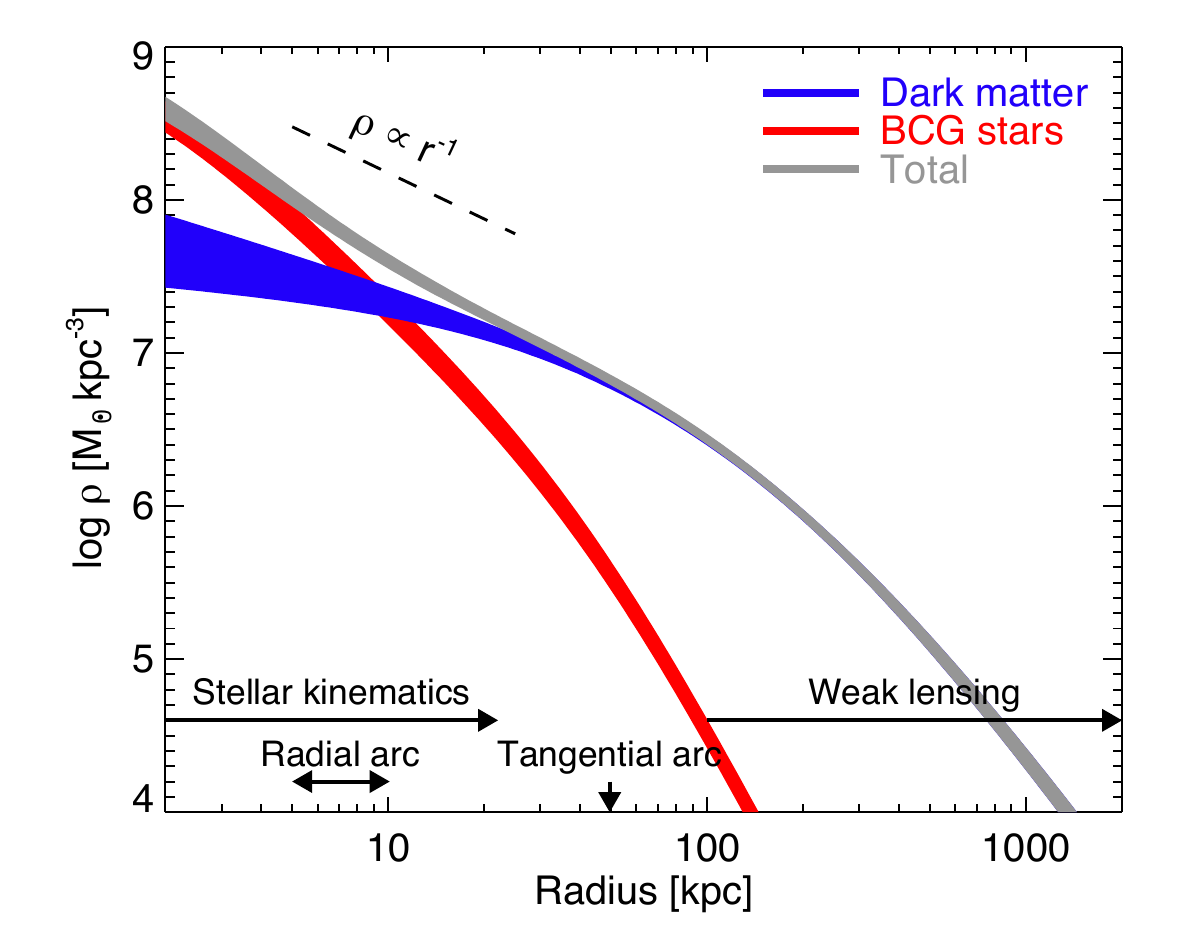,width=0.6\textwidth}}
\caption{Probing how dark matter is distributed on small scale via strong gravitational lensing in the core of a massive cluster. The left panel shows a Hubble Space Telescope image of the centre of the rich cluster Abell 383. A radial arc (images 1A, 1B) is visible close to the massive central galaxy (whose image has been subtracted to improve the visibility of faint features). The position and redshift ($z$=1.01) of this radial arc encodes the density profile of material close to the cluster centre, while the tangential arc (images 2A-C) measure the total enclosed mass. The red curve shows the critical line at this redshift. In the standard cold dark matter model, radial density profiles (right panel) are predicted to be very steep ($\rho\propto\,r^{-1}$ - black dashed line), in conflict with the inferred dark matter density profile from observations (blue curve). Such differences may be explained either by physical processes occurring in such dense environments (e.g. supernovae explosions) which redistribute both the baryonic and dark matter or, more interestingly, by the suggestion that dark matter is `self-interacting'  (after \cite{New++13}).}
{\label{fig:Newman}}
\end{figure*}

Clusters of galaxies acting as a gravitational lens, provide a unique
opportunity to shed light on dark matter in a particularly clean
manner. When a source is very well aligned with the cluster centre,
two of the multiple images can form a {\it radial arc}.  The position
of the radial arc is a direct measurement of the slope of the enclosed
mass density profile close to the centre of the cluster. Incorporating
other evidence, it is possible to derive the dark matter density
profile of the cluster with unprecendented precision
\citep{New++13}. The results of this work also show that in some cases the
dark matter density profile is significantly flatter than predicted in
simulations based on the standard cold dark matter model
(Figure~\ref{fig:Newman}).  Significant theoretical efforts
\cite{Elz++04,Nip++04,Lap++12,L+W14,Sch++14} are currently underway to
clarify whether this disagreement is due to a poor understanding of
the physics of how baryons interact with dark matter or, more
dramatically, whether dark matter has some property that is not yet
understood. One appealing solution suggests that dark matter can
interact with itself \cite{Roc++13,Kap++14}. In dense regions, such as
the centre of a massive cluster, such {\it self-interacting dark matter}
would naturally flatten the predicted cusps in agreement with
observations. Interestingly, lensing observations of colliding
clusters already set an upper limit to amount of self interaction
\cite{Clo++06,Ran++08}, so that the range of astrophysically interesting 
interaction strength allowed by lensing obervation is limited.

\subsection{`Natural' telescopes}

With extraordinary vision, in 1937 Fritz Zwicky \cite{Zwi37} suggested
that clusters of galaxies could be used as `natural telescopesÕ to
search for magnified images of very distant galaxies, thereby
extending the reach of our existing telescopes'. In the past decade
this has become a very effective way to locate and understand the
properties of the earliest galaxies seen when the Universe was less
than 10\% of its current age. A few hundred millions years after the
Big Bang, the first stellar systems emerged, as yet unpolluted by
nuclear enrichment given only light elements were synthesised in the
beginning. These systems were very hot and emitted copious amounts of
ultraviolet radiation capable of photo-ionising the hydrogen in deep
space. The arrival of the first galaxies, termed {\it cosmic dawn} is
thought to be responsible for this {\it cosmic reionisation} \cite{Rob10}.

\begin{figure*}
\centerline{\psfig{file=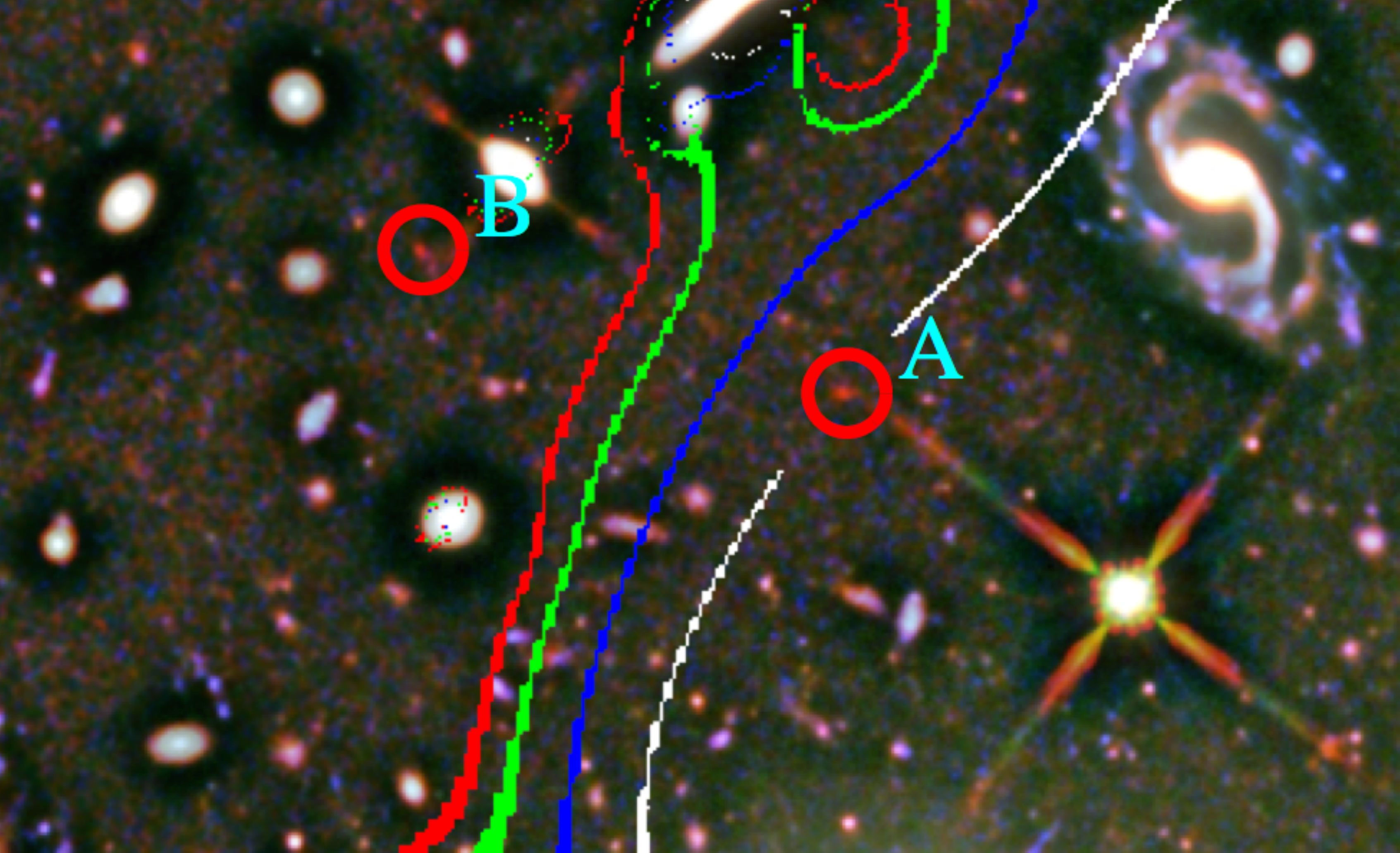, width=0.6\textwidth}}
\caption{Identification of one of the most distant known galaxies via the magnification of the gravitational lensing phenomenon The red colors of the multiply-imaged system (A and B) together with the symmetry of the position around the `critical line' of high magnification for a distant source indicates  that it is likely at $z\sim10$, seen when the Universe was barely 5\% of its present age. The dependence of the location of the critical line on source redshift is indicated by the red, green, blue and white lines which are for sources at redshifts 10, 3.6, 2.0 and 1.3, respectively (after \cite{Zitrin14}).}
\label{fig:Zitrin}
\end{figure*}

The Hubble Space Telescope has given us our first glimpse of this last
remaining frontier of cosmic history, but individual galaxies seen
from this remote past are too faint for detailed study. All we can do
at present is a census of their abundance and luminosity
distribution. To understand whether, for example, the typical stars in
these galaxies are pristine requires spectroscopic investigations
which will be challenging, even with the next generation of ground and
space-based facilities. This is precisely where gravitational lensing
can help. A foreground cluster of galaxies presents a large
cross-section to this background population so the likelihood of
magnified images is high (Figure~\ref{fig:a2744}). On the other hand,
the distribution of mass in a cluster is less regular than in a single
galaxy, so careful modeling is necessary to quantify the boost in
signal. Some of the most distant galaxies known have been located by
searching close to the so-called critical lines of massive clusters
where magnifications of $\times$20-30 are possible
\cite{Yee+96,Ell++01,Kne++04,Bra++09,Coe++13}; these systems would
otherwise not have been detected. The Hubble Space Telescope is now
undertaking a systematic deep survey of six such foreground clusters
with its near-infrared camera (Figure~\ref{fig:a2744}). This programme,
termed the Hubble Frontier Fields\footnote{\tt
http://www.stsci.edu/hst/campaigns/frontier-fields/}, has already
revealed the most distant galaxies known
\cite[][Figure~\ref{fig:Zitrin}]{Zitrin14}, and promises to deliver
many examples of highly-magnified systems for subsequent study
e.g. with the James Webb Space Telescope, a near-infrared large
aperture space telescope due for launch in 2018.

Clusters not only magnify sources in their integrated brightness,
rendering them more easily visible with our telescopes, but lensing
also enlarges the angular size of a distant source making it easier to
determine its internal properties. The most distant galaxies are
physically very small -- about 10 times less so than our Milky Way --
and resolving them is a challenge for both Hubble Space Telescope and
large ground-based telescopes equipped with adaptive optics (AO) -- a
technique that corrects for atmospheric blurring.  However the
combination of high resolution imaging from HST or AO {\it and}
gravitational magnification offers spectacular opportunities. A
distant galaxy at a redshift of 3 is typically only 0.2-0.3 arcsec
across yet, when magnified by a factor of $\times$30, it is possible
to secure spectroscopic data point-by-point across the enlarged images
of representative examples. Such studies of lensed $z\simeq$2-3
galaxies observed during a time of peak activity in galaxy assembly
demonstrate that many have primitive rotating disks \cite{Sta++08},
\cite{Wuy++14}, and that there is a range of chemical
gradients in their gaseous composition \cite{Jon++13,Jon++14}.

\section{Weak lensing and Cosmic Shear}
\label{sec:weak}

Strong lensing signatures are usually straightforward to recognise in
astronomical images via the presence of multiply-imaged sources, often
with highly distorted shapes. Weak lensing, which affects all photons
travelling in the universe is much harder to recognize. The principal
signal is a small distortion in the shape of a background galaxy which
depends on the curvature (or second derivative) of the foreground
gravitational potential.  In an idealised case, the observer sees the
background source slightly stretched (or sheared) tangentially around
a circle whose center is the lensing structure
(Figure~\ref{fig:sketch}). Unfortunately, this weak lensing signal is too
feeble to be inferred for a single background source and, of course,
we would need to assume the true shape and orientation of a source to
measure it quantitatively. However, the presence of foreground
structures can still be inferred by statistically averaging the
distorted shapes of many background galaxies in a given direction
assuming the sources are, overall, randomly oriented. Since it is so
pervasive, weak lensing offers the prospect of making maps of the
otherwise invisible dark matter everywhere in the sky, on scales much
larger than galaxies or clusters of galaxies. Consider the following
remarkable fact: the night sky does not present us with a faithful
picture of the Universe!  In all directions light rays from distant
galaxies are being subtly deflected as they traverse the cosmos by low
density dark matter structures -- a phenomenon we call {\it cosmic shear}.

Weak gravitational lensing holds enormous promise in observational
cosmology as the technique, properly employed, can reveal the
distribution of dark matter independently of any assumptions about its
nature. However, the technical challenges are formidable. Foremost the
signal arising from large scale structure is small -- amounting to a
change in the ellipticity of a faint distant galaxy of only a few
percent. Secondly, as a statistical technique, a high surface density
of measurable galaxies must be secured in order to gather sufficient
signal, so deep imaging is essential. Finally, as the Earth's
atmosphere smears the shapes of faint galaxies, painstaking
corrections must be made to recover the cosmological signal.

\begin{figure*}
\centerline{
\psfig{file=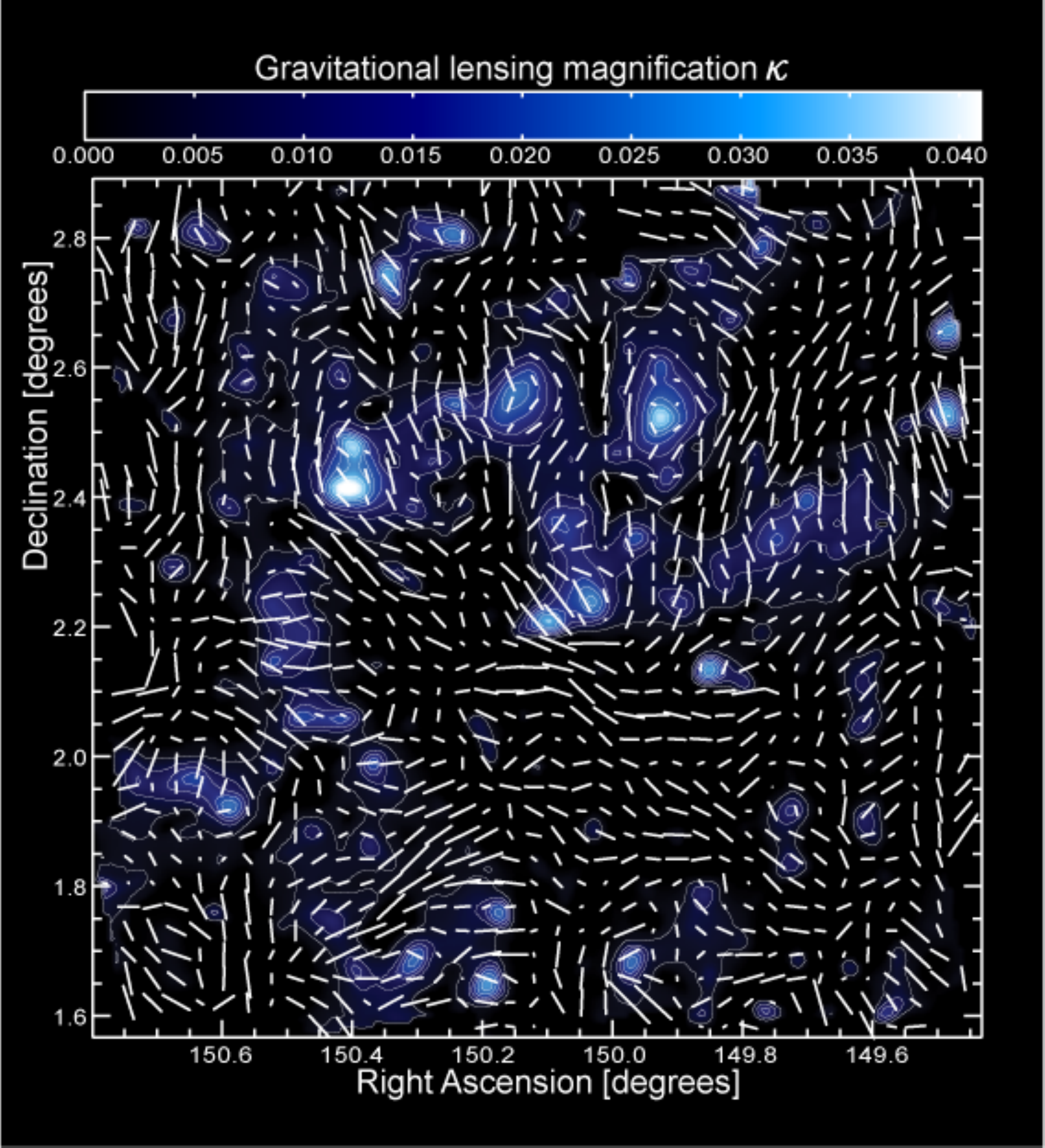,width=0.8\textwidth}}
\caption{Mapping dark matter using weak gravitational lensing. The vector tick marks represent the mean alignments of faint galaxies across a 2 square degree field imaged by the Hubble Space Telescope. The pattern can be used to make a projected 2-D map of the dark matter along the line of sight. Where the pattern swirls tangentially around a region, it indicates a mass concentration in that direction; where the pattern is radially outward, an under density. By slicing such day matter maps according to look-back time, the growth of structure in the Universe can be traced. This is a powerful measure of the competition between the positive gravitational influence of dark matter and the repulsive effect of dark energy (after \cite{Mas+07}). \label{fig:massey}}
\end{figure*}

The first claimed detection of a weak lensing signal was by Tyson and
collaborators \cite{TWV90} in the field of a rich cluster. But the
techniques for robustly analyzing the pattern of distortions of
background galaxies and inverting these to map the foreground dark
matter were developed by Kaiser and others soon after
\cite{Kai92,KSB95}. Detections of weak lensing from the large scale
distribution of dark matter along random sightlines, that is ``cosmic
shear'', were not announced until 2000
(\cite{Bac00}\cite{vWa00}\cite{Wit00}).  These early papers analyzed
the strength of the signal to constrain the amount of dark matter per
unit volume, confirming independently values from other methods. Later
papers used the techniques to produce maps of the projected dark
matter distribution
\cite[Figure~\ref{fig:massey}][]{Mas+07} \cite[see also][]{G+S07}. 
These maps can be compared to that of the light in the
same direction as revealed by visible galaxies and X-ray emitting clusters. To
first order there is a reassuring similarity, indicative of the fact
that dark matter acts as the gravitational framework (or scaffolding)
for the normal baryonic material.

Ambitious ground-based surveys are now being undertaken with
telescopes equipped with panoramic camera with a goal not only of
charting the distribution of dark matter in two dimensions but also
tracing its clustering with time. Structure formation develops
according to two basic forces - the attractive force of gravity and
the repulsive effect of dark energy.  The 3.6m Canada France Hawaii
Telescope has undertaken a survey of 154 deg$^2$ in five photometric
bands with the MegaCam 1 deg$^2$ CCD imager \cite{Hey12}. The
European Southern Observatory's 2.6m VST is undertaking the Kilo
Degree Survey (KIDS) - a 1500 deg$^2$ survey in four photometric bands
with a similar camera called OmegaCam 
\cite{deJ++13}. The Pan-STARRS project at the University of Hawaii
plans a series of telescope upgrades in order to chart the distribution of
dark matter. The Dark Energy Survey has just began and in the next 5 years it will image
5000 deg$^2$ of sky in 5 photometric bands using a 520 megapixel
camera on the 4m Blanco Telescope in Chile\footnote{
\rm http://www.darkenergysurvey.org/index.shtml}. Finally, the Subaru
Telescope is now equipped with the most impressive HyperSuprimeCam - a
870 megapixel CCD camera providing a 1.5 deg$^2$ field. This
instrument is likewise being used to undertake a 1200 deg$^2$
survey\footnote{\rm
http://www.naoj.org/Projects/HSC/surveyplan.html}. By virtue of the
larger aperture of the Subaru 8.2m, this will be the deepest weak
lensing survey in the next few years.

These surveys, collectively, aim to pinpoint the nature of dark energy
in two respects. Firstly, by comparing the rate at which structure
grows to the rate of cosmic expansion (for example from studies of
luminous supernovae or gravitational time delays), we gain insight
into whether dark energy is an illusion caused by a failure of
Einstein's theory of gravity on large scales. Surprising though it may
seem to challenge Einstein,who founded the very topic of this article,
one must remember that skeptics challenged Einstein when he dared to
insist that Newtonian gravity needed amending! Secondly, the data will
assist in determining whether dark energy is a constant or evolving
property of space. Right now the limited data imply that dark energy
is consistent with a constant energy density (sometimes called the
`cosmological constant', corresponding to $w=-1$ in the language of
the equation of state of dark energy (see Figure~\ref{fig:Suyu}) to a
precision of about 5\% over the past 8 billion years. However, these
upcoming surveys will significantly improve this constraint. Whatever
the outcome -- a new theory of gravity on large scales, verifying a
`cosmological constant' or discovering an evolving component of dark
energy -- new physical insight is guaranteed!

\section{Microlensing finds extrasolar planets}

\begin{figure*}
\centerline{
\psfig{file=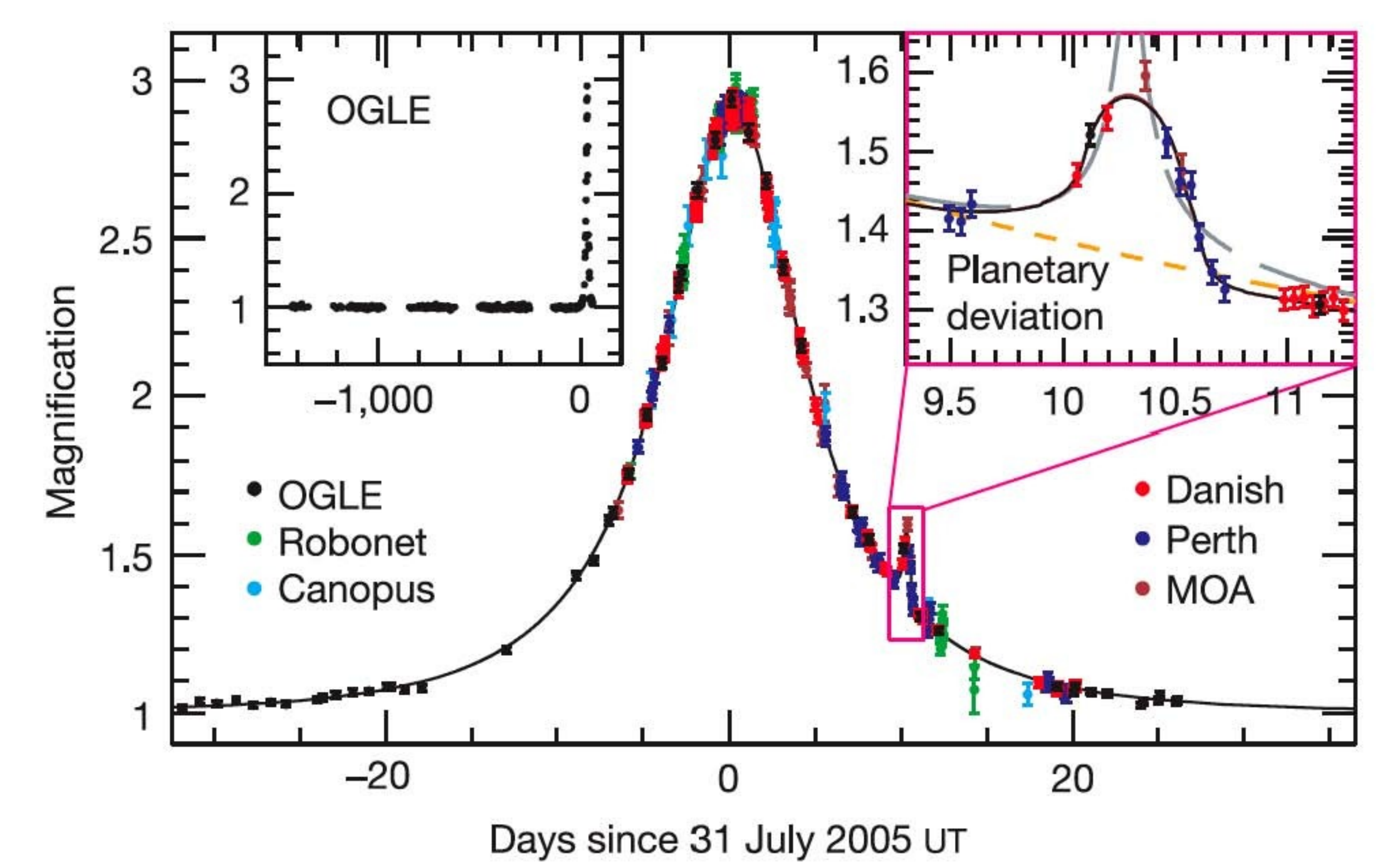, width=0.7\textwidth}}

\caption{Detection of a low mass exoplanet (OGLE 2005 BLG 390LB) via a small perturbation to the microlensing light curve. The light curve represents the time-dependent brightness of a background star in magnitudes (a logarithmic scale). As a foreground star crosses the path, it magnifies the background source. An associated planet in the foreground provides an additional boost over a shorter timescale. Microlensing can detect low mass planets further away from their host stars than other methods and thus promises a more complete inventory of their abundance in the Milky Way (after \cite{Bea++06}).}

\label{fig:Gould}
\end{figure*}

In 1936 Einstein was not convinced gravitational lensing would yield
observable returns because the probability that two stars are
sufficiently well-aligned so that one magnifies the other is very
low. But with panoramic imaging cameras, many tens of millions of
stars can be monitored efficiently. Together with the fact that there
can be relative motion between the source and lens, there is still the
likelihood of observing an effect. Microlensing - a term introduced by
Bohdan Paczynski - generally refers to the case where either the
source or both the source and lens are unresolved. Consequently, the
deflection and distortion of light from the background source cannot
be seen. The key signature is a temporal brightening of the combined
signal from source plus lens as the one passes in front of the
other. The timescale of the brightening can be anything from seconds
to years and the observed light curve gives information on the lens
mass, the relative distances and the motion of the lensing object
(assuming the background object is stationary). As microlensing is a
transient phenomenon, an effective survey strategy is to monitor a
dense stellar field repeatedly, searching for that rare occasion when
an individual star increases its brightness.  Of course, complicating
such searches is the fact that many stars are genuinely variable in
their output. Once a likely event has been triggered, it can be
monitored more intensively to see if the light curve is of the form
expected for microlensing.

Microlensing has had a major impact in astronomy in two areas, both
involving monitoring of tens of millions of stars in the Milky Way or
nearby Magellanic Clouds: (i) the search for dark matter in the
Galactic halo in the form of compact objects of moderate mass (~0.1
solar masses or less) and (ii) locating and assessing the abundance of
extrasolar planets down to as low as the mass of the Earth. In the
case where the lens comprises a star with an orbiting planet, the
light curve deviates from that expected for a single lens. Such a
signal was first observed in 2004 by Bond and colleagues \cite{Bon04}
leading to the detection of a 1.5 Jupiter mass planet.
Figure~\ref{fig:Gould} shows one example where a 5.5 Earth mass planet
was detected via a perturbation to the light curve of a microlensing
event. More recently a planet of two Earth masses has been detected in
an orbit similar in radius to that of the Earth around the Sun
\cite{Gou++14}.

Because it is a geometric technique, microlensing probes planet in the
cold outer regions, far from their host stars. By contrast, other
techniques for finding exoplanets, such as monitoring the radial
velocity of the host star or searching for transits, are sensitive to
planets much closer to their host stars. Moreover, in sharp contrast
to other methods, the signal does not necessarily decline as the
planet mass decreases. Microlensing has, like the other aspects of
lensing reviewed above, not yet been thoroughly exploited. It offers
the exciting prospect of a more complete inventory of planets around
stars in the Milky Way.

\section{The future: gravitational lensing in the next decade}

It is remarkable to consider that, as recently as 1970, gravitational
lensing was largely considered as a mathematical curiosity with no
practical purpose, neglected by the observational community. The only
observational progress was improved precision in measurements of the
solar deflection initiated in 1919 and in the Shapiro time delay \cite{Wil06}.

Today, there is a thriving community of scientists working on the
subject, so much that it has become impossible to do justice to the
entire field in a single review, but we had to focus on a few
highlights. This young and energetic community is not only making
great strides in answering profound scientific questions, but it is
also pushing the boundaries with new computational, statistical, and
mathematical tools \cite{Kee10}, as well as novel ways of conducting
collective science.  Teams of postdoctoral astronomers compare their
mass models for the clusters being imaged by HST for the Frontier
Field program at dedicated science workshops, while others participate
in blind tests of their weak lensing analysis codes by applying them
to simulated datasets \cite{Man++14}, or in blind tests of the
recovery of gravitational time delays \cite{Lia++14}.

Significant observational surveys are underway or planned, like the Large Synoptic Survey 
Telescope (LSST\footnote{\rm http://www.lsst.org/lsst/}). With dedicated state-of-the-art imaging
cameras to chart large areas of the sky in depth and through a range
of colour filters these surveys trace weak lensing signals at various
cosmic distances in order to map the evolving dark matter distribution
(see \S4) and learn about dark energy \cite{DESC12}. The same surveys
will be used to locate large numbers of strong lenses for detailed
studies using methods such as those described in \S2, when combined
with follow-up high resolution information from the James Webb Space
Telescope \cite{Gar++06} or the next generation of adaptive optics
systems on large and extremely large telescopes. The commissioning of
the Atacama Large Millimiter Array (ALMA) and the planned Square
Kilometer Array\footnote{\rm https://www.skatelescope.org/} guarantee
that radio wavelengths will also play an important role in the
scientific exploitation of the gravitational lensing effect
\cite{Koo++09b,Neg++10,Hez++13,Hez++14}.

The successful progress with these ground-based surveys, together with
evident advantage of high angular resolution data revealed by HST, has
given the international community the necessary impetus to plan two
ambitious satellite missions whose science case builds on the
cosmological opportunities afforded by gravitational lensing. The
European Space Agency's {\it Euclid} mission\footnote{\rm
http://www.euclid-ec.org/} is a 1.2 metre telescope with a 0.7 degree
diameter field of view which will undertake a survey of 15,000 square
degrees when it is launched in 2020. A high priority in NASA's future
programme is an even more ambitious mission currently called
WFIRST-AFTA\footnote{\rm http://wfirst.gsfc.nasa.gov/about/} which
will be based on a 2.4 metre telescope `inherited' from the US
National Reconaissance Office. The detailed science payload for this
mission is still being determined but panoramic imaging to limits even
deeper than achievable with {\it Euclid} are likely.

Of course, as in any relatively new field, with the giant leaps in
technology, there is also the possibility of discovering new hiterto
exotic phenomena, like lensing by cosmic strings \cite{Mor++10} or
lensing in the strong gravity field of a black hole \cite{B+M12,Gen14}.

Nearly one hundred years after Einstein's remarkable realization that
he could solve Newton's dilemma of how the gravitational force acts a
distance, one wonders what he would make of both the mysterious
composition of the Universe, dominated by dark matter and dark energy,
as well as how the phenomenon he predicted has now taken a central
role in making progress. Einstein was not afraid to admit his previous
errors of judgement on several occasions, for example in initially
disputing the expanding Universe. We suspect therefore that he would
be an enthusiastic promoter of gravitational lensing, as are we!

\section*{Acknowledgements}

TT acknowledges support by National Areonautics and Space
Administration (NASA), the National Science Foundation (NSF), and the
Packard Foundation through a Packard Research Fellowship. TT thanks
the American Academy in Rome and the Observatory of Monteporzio Catone
for their kind hospitality during the writing of this manuscript.


\label{refs}


\clearpage
\section*{About the Authors}

\begin{figure*}
\centerline{
\hbox{
\psfig{file=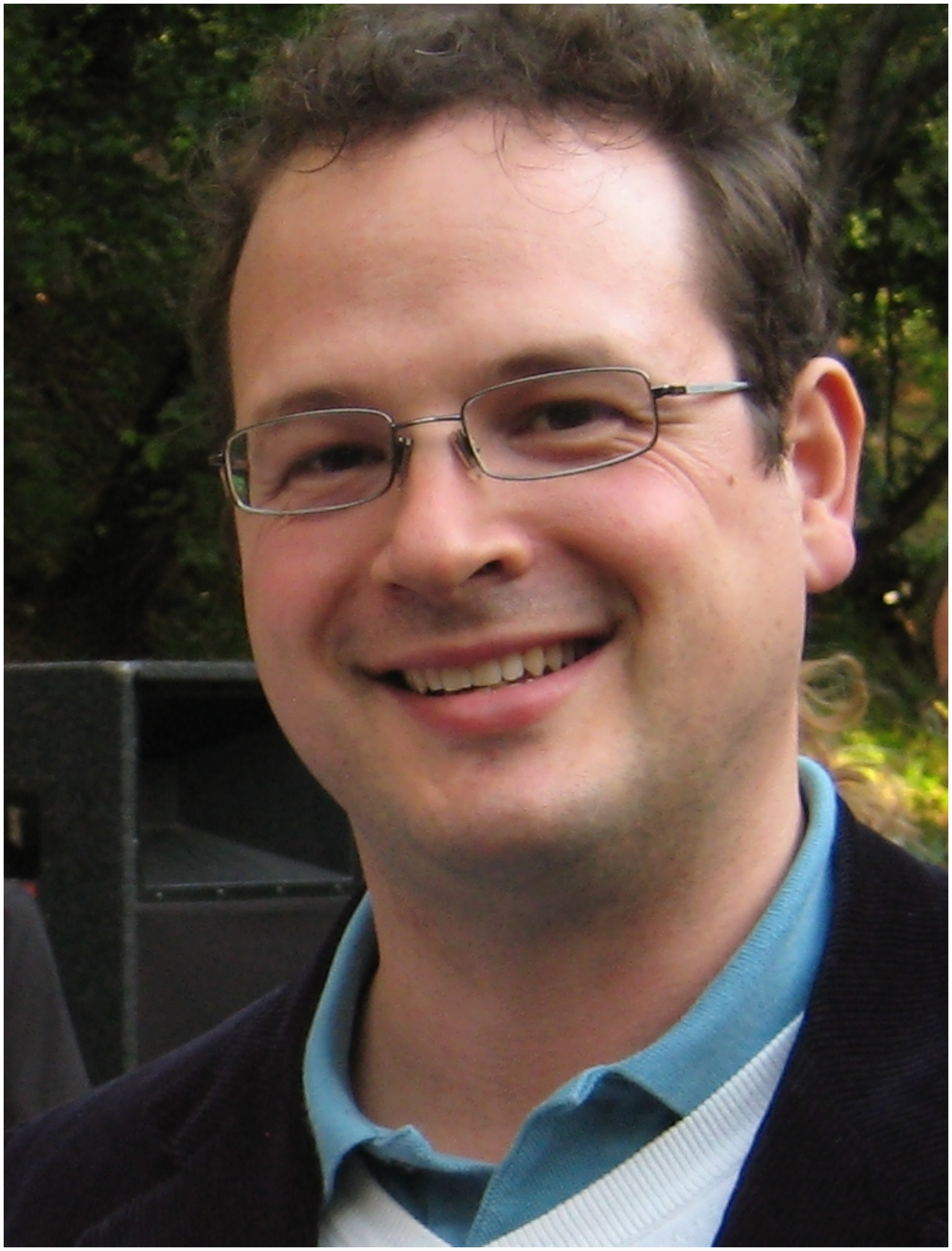,width=0.4\textwidth}
\psfig{file=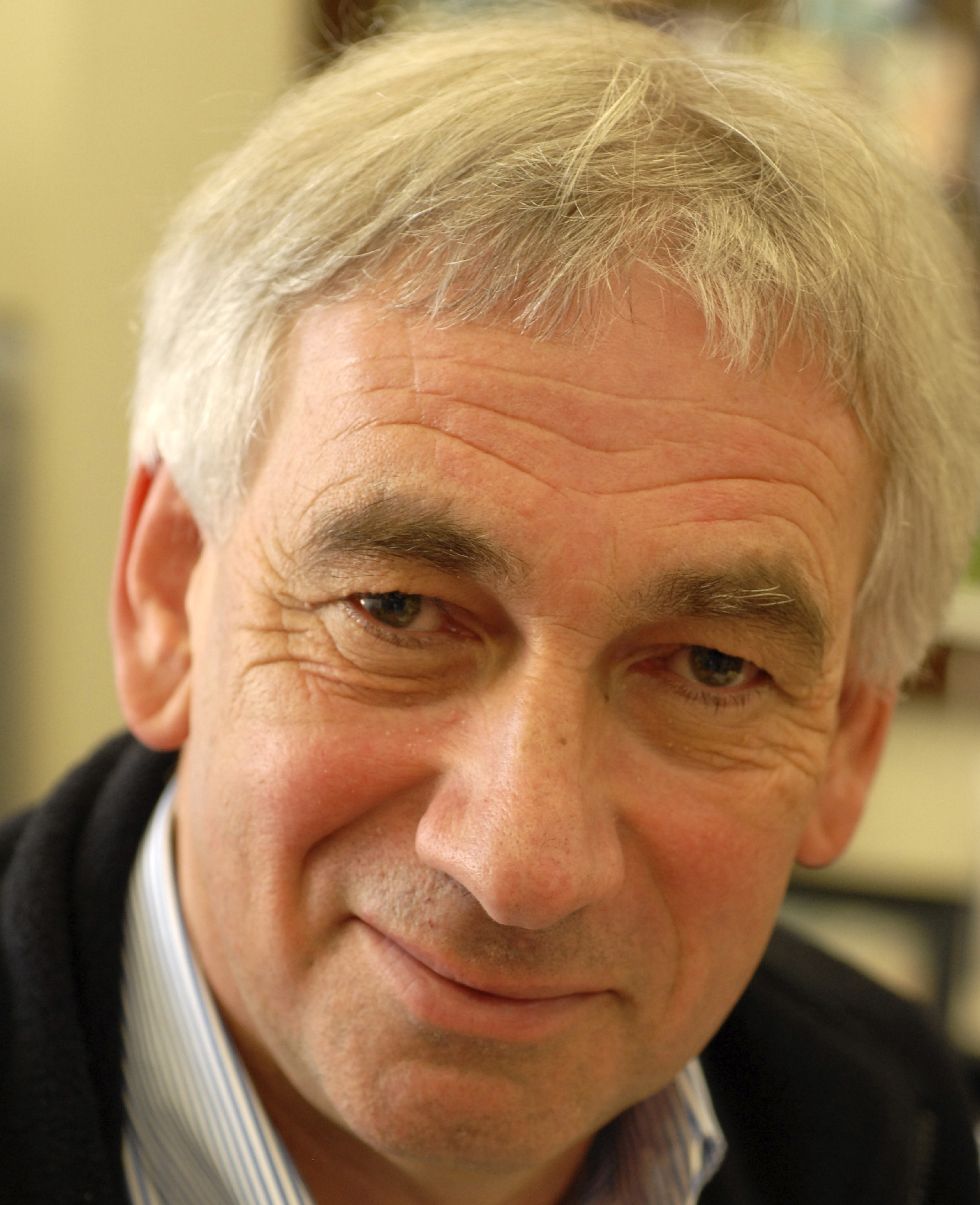,width=0.4\textwidth}}}
\end{figure*}

Tommaso Treu (left)  is currently Professor of Physics and Astronomy 
at the University of California, Los Angeles (UCLA) having recently 
transferred from UC Santa Barbara. Treu was educated at the University 
of Pisa and the Scuola Normale before undertaking postdoctoral
research at Caltech and UCLA. Treu has worked on numerous
projects in the areas of galaxy evolution, stellar dynamics and
gravitational lensing and published several reviews on these
topics.

\smallskip

\noindent Richard Ellis (right) is currently the Steele Professor of Astronomy at the
California Institute of Technology. He has held professorial positions
previously at the Universities of Durham, Oxford and Cambridge.
Ellis works in observational cosmology and galaxy evolution and
has played a key role in promoting the Thirty Meter Telescope
now under construction on Mauna Kea, Hawaii

\end{document}